# Entangling Dipole-Dipole Interactions and Quantum Logic in Optical Lattices

Gavin K. Brennen and Ivan H. Deutsch

*Center for Advanced Studies, Department of Physics and Astronomy,*

*University of New Mexico,  Albuquerque, New Mexico 87131*

Poul S. Jessen

*Optical Sciences Center,*

*University of Arizona, Tucson, Arizona 85721*

## Abstract

We study a means of creating multiparticle entanglement of neutral atoms using pairwise controlled dipole-dipole interactions in a three dimensional optical lattice. For tightly trapped atoms the dipolar interaction energy can be much larger than the photon scattering rate, and substantial coherent evolution of the two-atom state can be achieved before decoherence occurs. Excitation of the dipoles can be made conditional on the atomic states, allowing for deterministic generation of entanglement.  We derive selection rules and a figure-of-merit for the dipole-dipole interaction matrix elements, for alkali atoms with hyperfine structure and trapped in well localized center of mass states.  Different protocols are presented for implementing two-qubits quantum logic gates such as the controlled-phase and swap gate.  We analyze the fidelity of our gate designs, imperfect due to decoherence from cooperative spontaneous emission and coherent couplings outside the logical basis.  Outlines for extending our model to include the full molecular interactions potentials are discussed.

# I. INTRODUCTION

The ability to coherently manipulate multiparticle entanglement represents the ultimate quantum control of a physical system and opens the door to a wide variety of fundamental studies and applications, ranging from improvements in precision measurement [1] to quantum simulation [2,3] and quantum computation [4]. Several physical realizations have been proposed in quantum optics, including ion traps [5] and cavity QED [6], and "engineered" entanglement has been demonstrated in the laboratory for both of these systems [7, 8]. Entangling unitary transformations have been implemented in liquid-state NMR on pseudo-pure states of nuclear spins in small organic molecules [9], though true entanglement has not yet been produced in these thermal samples [10]. Proposals have been made also for a number of condensed-matter implementations, such as quantum dots [11], SQUIDS [12], and coupled spin-resonance of dopants in a silicon lattice [13]. All these implementations must contend with the conflict inherent to open quantum systems. A quantum computer must provide strong coherent coupling between the qubits and the external driving fields which run the algorithms, while shielding the qubits from the noisy environment that leads to decoherence. Moreover, many algorithms will require huge numbers of qubits, so the entangling mechanism must not degrade as the size of the system increases (i. e. it should be scalable).

Recently we identified a new method for producing multiparticle entangled states using cold trapped neutral atoms in optical lattices [14] (see also [15-17]). This system offers several advantages for quantum information processing. Generally neutral atoms in their electronic ground state couple extremely weakly to the environment. This also implies a weak coupling between atoms. Under special circumstances, however, interatomic couplings can be created on demand through external fields which excite two-atom resonances, such as those arising from electric dipole-dipole interactions [14], ground state collisions [15], or real photon exchange [16,17]. The ability to turn interactions "on" and "off" suppresses coupling to the decohering environment and the spread of errors among qubits. In our proposal this is achieved through external fields and controllable wave function overlap. The challenge to implementing multiparticle entanglement with neutrals is to design and carry out precise quantum control of the atoms, both in terms of their internal and external degrees of freedom, so that these couplings are turned on and off in a coherent manner and with high fidelity.

We can characterize the ability of the system to perform entangling unitary operations via a figure of merit that measures the ratio of the coherent interaction energy of two qubits to their collective decoherence rate. A simple scaling argument shows the plausibility of using dipole-dipole interactions between neutrals for coherent quantum control. Given two atoms separated at a distance $r$ in the near field, the interaction energy scales as $V_{dd} \sim \langle \mathbf{d} \rangle^2 / r^3$, where $\langle \mathbf{d} \rangle$ is the mean induced dipole moment per atom. When a resonant field is tuned to produce dipole-dipole coupling, a decohering channel becomes available because the induced dipoles can spontaneously emit photons. However, this process is bounded above by the Dicke-superradiance cooperative decay rate, which is equal to twice the single atom scattering rate $\sim k_L^3 \langle \mathbf{d} \rangle^2 / \hbar$, where $k_L$ is the wave number of the driving laser. We define a figure of merit for this interaction as $= V_{dd} / \hbar \quad \sim (k_L r)^{-3}$. Thus, if the atoms are tightly confined relative to the optical wavelength, the dipole-dipole level shift can be much larger than the scattering rate, and the coherent interaction can cause the atomic wave functions to acquire a substantial phase shift before decoherence occurs. This separation of time scales is the central feature that makes coherent quantum logic possible.

In order to achieve the tight localization required for coherent interaction, the atoms must be trapped and cooled. We explore the use of three dimensional optical lattices as a trap in which to create entangled states of atoms. When the light field forming the optical lattice is both intense and detuned far from atomic resonance, $|\Delta| = |\omega_L - \omega_0| \gg \gamma$, where $\gamma$ is the single atom resonance linewidth, then the mean dipole moment will be very small, making the scattering rate negligible, while the large field amplitude maintains a substantial trapping potential. Pure quantum states can be prepared by cooling to the vibrational ground state of the optical lattice using Raman sideband cooling [18], or perhaps by loading the lattice with a precooled sample of Bose-condensed atoms [19]. Both internal and external degrees of freedom can be further manipulated through the application of coherent laser pulses as in ion traps [20]. The optical lattice is thus seen to offer two key resources: a high degree of quantum control combined with an architecture that is naturally scalable to operate in parallel on many atoms [21]. Other possibilities for trapping and manipulating ultracold atoms exist, including miniature magnetic traps [22], but these will not be addressed here.

In this paper we consider controlled entanglement via resonantly induced dipole-dipole interactions between atoms in tightly confining traps. Our goal here is to establish some general features, applicable in a variety of settings. Experimental details will generally be omitted except to give order of magnitude estimates where appropriate. We begin with a discussion in Sec. II of the dipole-dipole interaction, establishing the selection rules for transitions between internal and external states for two atoms in confining traps. In Sec. III we demonstrate the flexibility available in this systems for implementing quantum logic by presenting three different two-qubit logic protocols and estimate the figure of merit for each due to spontaneous emission. A major deficit in the present analysis is the omission of inelastic collision channels from the model [23]. Such a task has considerable challenges, especially when one includes the complex internal structure of the atoms in the molecular potentials [24]. In Sec. IV we give an outlook toward this and other future research, and summarize our main results.

## II THE DIPOLE-DIPOLE INTERACTION

The dipole-dipole interaction depends both on the internal electronic states of the atoms, which set the tensor nature of the interaction, and the external motional states, which determine the relative coordinate probability distribution of the dipoles. We consider a system of two atoms trapped in harmonic wells, interacting coherently with a classical field and with each other via the dipole-dipole interaction. Decoherence may occur via cooperative spontaneous emission. We seek expressions for the interaction matrix elements and the resulting selection rules.

Consider two alkali atoms with nuclear spin $I$ and center of mass positions $\mathbf{r}_1$, $\mathbf{r}_2$, and excited on the D2 transition $\left| S_{1/2}(F) \right\rangle \to \left| P_{3/2}(F') \right\rangle$, where $F$ and $F'$ belong to the ground and excited state hyperfine manifolds. The atoms interact with the vacuum field and a classical monochromatic laser field $\mathbf{E} = \mathrm{Re}\left( E_L \vec{\varepsilon}_L(\mathbf{x}) e^{-i\omega_L t} \right)$, with amplitude $E_L$ and local polarization and phase given by $\vec{\varepsilon}_L(\mathbf{x})$. After tracing over the vacuum modes in the usual Born-Markov and rotating wave approximations, one obtains the effective Hamiltonian for the atom-laser interaction, together with a dipole-dipole interaction between atoms [25],

$$H_{AL} = -\hbar\left( \Delta + i\frac{\Gamma}{2} \right)\left( \mathbf{D}_1^{\dagger} \cdot \mathbf{D}_1 + \mathbf{D}_2^{\dagger} \cdot \mathbf{D}_2 \right) - \frac{\hbar}{2}\left( \mathbf{D}_1^{\dagger} \cdot \vec{\varepsilon}_L(\mathbf{r}_1) + \mathbf{D}_2^{\dagger} \cdot \vec{\varepsilon}_L(\mathbf{r}_2) + h.c. \right), \quad (1a)$$

$$H_{dd} = V_{dd} - i\frac{\hbar\Gamma_{dd}}{2} = -\frac{\hbar\Gamma}{2}\left( \mathbf{D}_2^{\dagger} \cdot \overset{\leftrightarrow}{\mathbf{T}}(k_L r) \cdot \mathbf{D}_1 + \mathbf{D}_1^{\dagger} \cdot \overset{\leftrightarrow}{\mathbf{T}}(k_L r) \cdot \mathbf{D}_2 \right). \quad (1b)$$

Here $\Gamma$ is the spontaneous emission rate for the $\left| P_{3/2} \right\rangle \to \left| S_{1/2} \right\rangle$ transition and $\Delta$ is the laser field detuning, assumed large compared to the excited stated hyperfine splitting (though not necessarily large with respect to the ground state splitting). The relevant Rabi frequency is given by $\Omega = \left\langle J' = 3/2 \| d \| J = 1/2 \right\rangle E_L / \hbar$, having used the Condon and Shortley [26] convention for reduced dipole matrix elements. The dimensionless dipole raising operator associated with absorption of a photon is defined as

$$\mathbf{D}^{\dagger}_{F} = \frac{P_F\, \mathbf{d}\, P_F}{\langle J \,\|d\|\, J\rangle} = \sum_{F} \gamma_{FF} \sum_{q=-1}^{1} \sum_{M_F} \mathbf{e}_q^{*}\, c^{F,1,F}_{M_F,q,M_F+q} \big|F\,,M_F+q\big\rangle\big\langle F,M_F\big| \,, \tag{2a}$$

$$\gamma_{FF} = \sqrt{(2J+1)(2F+1)} \begin{Bmatrix} F & I & J \\ J & I & F \end{Bmatrix} \,, \tag{2b}$$

where $P_{F,F}$ are projectors on the excited and ground manifolds, $\mathbf{e}_q$ are the spherical basis vectors, $c^{F,1,F}_{M_F,q,M_F+q}$ is the Clebsch-Gordan coefficient for the electric dipole transition $\big|F,M_F\big\rangle \rightarrow \big|F,M_F+q\big\rangle$, and $\gamma_{FF}$ is the relative oscillator strength of the transition as set by the 6$j$-symbol. The second rank tensor, $\vec{\mathbf{T}} = \vec{\mathbf{f}} + i\vec{\mathbf{g}}$, describes the strength of the two-atom interaction as a fun tion of atomic separation $r = \big|\mathbf{r}_1 - \mathbf{r}_2\big|$,

$$\vec{\mathbf{f}}(k_L r) = \frac{3}{2}\left[ \left(\vec{\mathbf{1}} - \hat{\mathbf{r}}\hat{\mathbf{r}}\right)\frac{\cos(k_L r)}{k_L r} - \left(\vec{\mathbf{1}} - 3\hat{\mathbf{r}}\hat{\mathbf{r}}\right)\left(\frac{\sin(k_L r)}{(k_L r)^2} + \frac{\cos(k_L r)}{(k_L r)^3}\right)\right] \,, \tag{3a}$$

$$\vec{\mathbf{g}}(k_L r) = \frac{3}{2}\left[ \left(\vec{\mathbf{1}} - \hat{\mathbf{r}}\hat{\mathbf{r}}\right)\frac{\sin(k_L r)}{k_L r} + \left(\vec{\mathbf{1}} - 3\hat{\mathbf{r}}\hat{\mathbf{r}}\right)\left(\frac{\cos(k_L r)}{(k_L r)^2} - \frac{\sin(k_L r)}{(k_L r)^3}\right)\right] \,. \tag{3b}$$

The Hermitian part of the effective interaction Hamiltonian, $V_{dd}$, determines the dipole-dipole energy level shift, whereas the anti-Hermitian part, $\gamma_{dd}$, gives rise to cooperative spontaneous emission, so that the total decay rate is given by the expectation value of $\gamma_{tot} = \gamma\left(\mathbf{D}_1^{\dagger}\cdot\mathbf{D}_1 + \mathbf{D}_2^{\dagger}\cdot\mathbf{D}_2\right) + \gamma_{dd}$. In the near field, taking the limit $k_L r \rightarrow 0$, one finds $V_{dd} \rightarrow \left(\mathbf{d}_1\cdot\mathbf{d}_2 - 3(\hat{\mathbf{r}}\cdot\mathbf{d}_1)(\hat{\mathbf{r}}\cdot\mathbf{d}_2)\right)/r^3$, the quasi-static dipole-dipole interaction, and $\gamma_{dd} \rightarrow \gamma\left(\mathbf{D}_1\cdot\mathbf{D}_2^{\dagger} + \mathbf{D}_2\cdot\mathbf{D}_1^{\dagger}\right)$, the Dicke super- (or sub) radiant interference term for in (or out of) phase dipoles. Because the level shift diverges for small $r$ whereas the cooperative emission remains finite, the time scales for coherent and incoherent interactions separate, providing a mechanism for controlled entanglement of the atoms.

In order to analyze our system we must choose an appropriate basis of states. There are two natural choices: the atomic and molecular bases. In the atomic case we consider product states of internal dynamics and center of mass motion. In the molecular case we consider eigenstates of the dipole-coupled two atom Hamiltonian $H = H_{A1} + H_{A2} + V_{dd}$ in the Born-Oppenheimer approximation. Both bases form a complete set of states and thus allow for a full description of the physics, though the transparency of the model may be greater with one or the other, depending of the nature of the problem. For low atomic densities the atomic basis is convenient (see for example [27]), whereas at high densities where collisions play a dominate role the molecular basis is more natural (see for example [23]). A simple example which makes explicit the relation between these descriptions for two 2-level atoms is reviewed in Appendix A (see also [28]). For realistic atoms, such as the alkalis treated here with both fine and hyperfine structure, the molecular basis has a very complex description [24]. In order to gain some intuition on our problem we will restrict our attention to the atomic basis, paying careful attention to approximations made. We will return to discuss the more general problem in Sec. IV.

Consider then a product state with the two atoms in the same internal state, $| \ \rangle = |_{int} \rangle_1 |_{ext} \rangle_1 \ | \ _{int} \rangle_2 |_{ext} \rangle_2$, each with its mean dipole moment vector oscillating along the spherical basis vector $\mathbf{e}_q$. Under this circumstance the figure of merit for coherent dipole-dipole level shift can be calculated from Eq. (1),

$$= \frac{\left\langle V_{dd} \right\rangle}{\left\langle \ _{tot} \right\rangle} = \frac{-\hbar}{2\hbar} \frac{\left\langle D_q^\dagger D_q \right\rangle_{int} \left\langle f_{qq} \right\rangle_{ext}}{\left\langle D_q^\dagger D_q \right\rangle_{int} \left(1 + \left\langle g_{qq} \right\rangle_{ext}\right)} = \frac{-\left\langle f_{qq} \right\rangle_{ext}}{2\left(1 + \left\langle g_{qq} \right\rangle_{ext}\right)}. \tag{4}$$

This factor depends only on *geometry*, the external states and the direction of polarization. It is independent of the strength of the dipole, since the same matrix element for the atoms' internal states appears both in the numerator and denominator. The average over the external state is carried out with respect to the relative coordinate probability density, having traced over the center of mass of the two-atom system.

We will focus here on weak excitation of the dipoles. Adiabatic elimination of the excited states follows from second order perturbation theory in the limit of small saturation of the atomic transitions. When the detuning is large compared to the excited state level shifts we can neglect the change of the level structure due to the dipole-dipole interaction, and consider saturation of the atomic levels, *independent of the external motional states*. In this regime there is a separation between the light-shift and the dipole-dipole potential energy as shown for the simple case of two 2-level atoms in Appendix A. For the case of alkali atoms the effective Hamiltonian on the ground state manifold is [27]

$$
\begin{aligned}
H_{dd} &= V_{dd} - i\frac{\hbar\gamma_{dd}}{2} \\
&= -s\frac{\hbar\gamma}{2}\sum_{q,q'=-1}^{1}\left(f_{qq'}+ig_{qq'}\right)\left(\left(\mathbf{D}_1{}^*{}_L(1)\right)\left(\mathbf{D}_1^\dagger{}_{q'}\right)\left(\mathbf{D}_2{}^*{}_{q'}\right)\left(\mathbf{D}_2^\dagger{}_L(2)\right)+h.c.\right),
\end{aligned}
\tag{5a}
$$

with saturation

$$
s = \frac{\Omega^2/2}{\Delta^2+(\gamma/2)^2} << 1.
\tag{5b}
$$

Here the interaction tensor is written in the spherical basis,

$$
\begin{aligned}
f_{qq'}(k_L r) &= \left(n_0(k_L r)\delta_{qq'}+(-1)^q n_2(k_L r)Y_2^{q'-q}(\theta,\phi)\sqrt{6\pi/5}\,c^{1,1,2}_{-q,q',-q'+q}\right), \\
g_{qq'}(k_L r) &= \left(j_0(k_L r)\delta_{qq'}+(-1)^q j_2(k_L r)Y_2^{q'-q}(\theta,\phi)\sqrt{6\pi/5}\,c^{1,1,2}_{-q,q',-q'+q}\right),
\end{aligned}
\tag{6}
$$

where $j_m (n_m)$ are the $m$th order spherical Bessel (Neumann) functions and $(\ ,\ )$ are the spherical angles of the relative coordinate $\mathbf{r}$. The zeroth order Bessel and Neumann functions account for retardation in the dipole-dipole interaction and will be neglected below.

Physically, Eq. (5) represents a four photon process: absorption of a laser photon by one atom followed by coherent exchange of the excitation between the atoms via a virtual photon emission and aborption, and finally stimulated emission of a laser photon returning both atoms to the ground state. Because the virtual photon can be emitted in any direction it is not an eigenstate of angular momentum with respect to the space-fixed quantization axis of the atoms. The quantum numbers $q$ and $q'$ represent two of the possible projections of its angular momentum on that axis. Examples of these fundamental photon exchange processes are shown in Fig. 1.

We are left to consider the geometry of the trapping potentials, resulting external coordinate wave functions, and the polarization of the oscillating dipoles. For deep traps we can approximate the motional states as harmonic oscillators. For the particular case of an isotropic trap, the spherical symmetry allows explicit evaluation of the interaction matrix elements. Consider two atoms in a common well, each described by a set of radial and angular momentum vibrational quantum numbers $|n,l,m\rangle$ [29], with energy $E_{nl} = 2n + l + 3/ $ , degeneracy $g_{nl} = (2n + l + 1)(2n + l + 2)/ 2$, and an internal state denoting one of the ground magnetic sublevels of a given hyperfine state $|F, M_F\rangle$. One can decompose the product state of the two isotropic harmonic oscillators into relative and center of mass states, which then can be used to find analytic expressions for the matrix elements with a general tensor coupling. Given dipoles excited with polarization $\mathbf{e}$ , we can evaluate the matrix element with respect to the internal degree of freedom,

$$\langle F, M_{F1}; n_1 l_1 m_1 | \quad \langle F, M_{F2}; n_2 l_2 m_2 | V_{dd} | F, M_{F1}; n_1 l_1 m_1 \rangle \quad | F, M_{F2}; n_2 l_2 m_2 \rangle =$$

$$\sum_{q, q'=-1}^{1} (-1)^q c_{-q\, q'\, -q+q'}^{1,1,2} \sqrt{6 / 5} \langle n_1 l_1 m_1 | \quad \langle n_2 l_2 m_2 | V(r) Y_2^{q'-q}(\ ,\ ) | n_1 l_1 m_1 \rangle \quad | n_2 l_2 m_2 \rangle$$

$$\left( c_{M_{F2}, \ , M_{F2}+}^{F,1,F} \ c_{M_{F2}+ \ ,-q \ ,M_{F2}}^{F,1,F} \ c_{M_{F1},q,M_{F1}+q}^{F,1,F} \ c_{M_{F1}+q,- \ ,M_{F1}}^{F,1,F} \ M_{F1}+q,M_{F1}+ \quad M_{F2}+ \ ,M_{F2}+q \right. +$$

$$\left. c_{M_{F1}, \ , M_{F1}+}^{F,1,F} \ c_{M_{F1}+ \ ,-q \ ,M_{F1}}^{F,1,F} \ c_{M_{F2},q,M_{F2}+q}^{F,1,F} \ c_{M_{F2}+q,- \ ,M_{F2}}^{F,1,F} \ M_{F1}+ \ ,M_{F1}+q \quad M_{F2}+q,M_{F2}+ \right)$$

(7)

where $V(r) = -s\hbar \quad n_2(k_L r)/2 \sim 1/^3$, having neglected the radiation term $n_0(k_L r) \sim 1/$ for

$\ll 1$. The Clebsch-Gordan coefficients of the four-photon process dictate selection rules for

the internal states,

$$M_{F1} + M_{F2} = (M_{F1} + M_{F2}) + (q - q').$$

(8)

We see that neither $M_{F1}, M_{F2}$, nor the total $M_{F1} + M_{F2}$ is a conserved quantity. The fact that

these are not good quantum numbers can be seen immediately from the form of the interaction

Hamiltonian, Eq. (1b), which is neither a scalar with respect to rotations by hyperfine operators

$\hat{\mathbf{F}}_1$, $\hat{\mathbf{F}}_2$, nor $\hat{\mathbf{F}}_1 + \hat{\mathbf{F}}_2$. Generally, internal angular momentum can be converted to rotational energy of the molecule if the atoms have multiply degenerate energy levels. Classically this is reflected in the fact that the dipole-dipole interaction is *not a central force*, and therefore the angular momentum of two classical dipoles about their center-of-mass is not a conserved quantity.

As an example of a fundamental processes which does not conserve internal quantum numbers, consider the specific case of a process for which $q = q'$. Though the total $M_F$ is a conserved quantity, due to degeneracies the internal angular momentum for *each atom* can change via exchange of virtual photons, as described above. An example of such a process is shown in Fig. 1b. Atoms A and B start in states $|F, M_F = 1\rangle$, $|F, M_F = -1\rangle$ respectively. Atom A absorbs a  -polarized laser photon, followed by exchange with B of a virtual photon whose angular momentum projection is $+$ in both absorption and emission, and finally stimulated emission of a laser photon. In the resulting dipole-dipole interaction, the atoms have both been transferred to internal state $|F, M_F = 0\rangle$. In general, for degenerate ground states, the two-atom state will evolve into an entangled state of internal and external degrees of freedom through interactions described by all allowed tensor components of $f_{qq}$. Techniques to conserve internal quantum numbers by breaking ground-state degeneracy are discussed in Sec. III.

To evaluate the external-state matrix element in Eq. (7) we first expand the uncoupled angular momentum basis for the motional states, $|l_1 m_1\rangle \, |l_2 m_2\rangle$, in terms of the coupled states $|\,,\mu\rangle$ for total angular momentum and projection $\mu$ according to the usual vector addition rules so that

$$\langle n_1 l_1 m_1| \, \langle n_2 l_2 m_2 |V(r)Y_2^{m_r}(\,,\,)| n_1 l_1 m_1\rangle \, |n_2 l_2 m_2\rangle =$$
$$\sum_{,\,\mu,\mu} c_{m_1,m_2,\mu}^{l_1,l_2,} c_{m_1,m_2,\mu}^{l_1,l_2,} \langle n_1 l_1 n_2 l_2; \,\,\mu |V(r)Y_2^{}(\,,\,)| n_1 l_1 n_2 l_2; \,\,\mu\rangle. \tag{9}$$

Borrowing a technique from nuclear physics due to Moshinsky [30] we express coupled isotropic harmonic oscillator wave functions for the two atoms in the product basis of center of mass oscillator states $|NL\rangle$ and relative coordinate oscillator states $|nl\rangle$ as summarized in Appendix B. The matrix element in Eq. (9) can then be written

$$\langle n_1 l_1 n_2 l_2; \,\,\mu |V(r)Y_2^{m_r}(\,,\,)| n_1 l_1 n_2 l_2; \,\,\mu\rangle$$
$$= \sum_{nNlLl} \langle n_1 l_1 n_2 l_2, \,| nl, NL, \,\rangle \langle n\,l\,, NL, \,| n_1 l_1 n_2 l_2, \,\rangle \times$$
$$(-1)^{L+\,+l} \sqrt{\frac{5(2\,+1)(2l+1)}{(4\,)}} c_{\mu\,,m_r,\mu}^{\,,2,} c_{0,0,0}^{l,2l,} \left\{ \begin{array}{ccc} l & & L \\ l & 2 & \end{array} \right\} \sum_{p=(l+l)/2}^{(l+l)/2+n+n} B(nl, n\,l\,, p) I_p(V(r)), \tag{10}$$

with restrictions on the quantum numbers to conserve the total mechanical energy,

$$E_{tot} = E_1 + E_2 = E_1' + E_2' = E_{CM} + E_{rel} = E_{CM} + E_{rel}' \qquad (11a)$$

$$n' = n + n_1 + n_2 - n_1' - n_2' + \left( l_1 + l_2 - l_1' - l_2' + l - l' \right) / 2. \qquad (11b)$$

In Eq. (10) $\langle n_1 l_1 \ n_2 l_2, \ | nl, NL, \rangle$ are Moshinsky brackets which are tabulated real coefficients found using recursion relations, $B(nl, n'l', p)$ are radial function expansion coefficients given in [31], and $I_p(V(r))$ are the Talmi integrals given in Appendix B. The Clebsch-Gordan coefficients in Eq. (9) impose the constraint $\mu = m_1 + m_2$, and $\mu' = m_1' + m_2'$, while Eq. (10) requires $\mu - \mu' = m_r$. Thus, the selection rule for a tensor coupling between isotropic harmonic oscillator states is

$$m_1' + m_2' + m_r = m_1 + m_2. \qquad (12)$$

This is the analog of Eq. (8), now with respect to the atomic motional states. The quantum number $m_r$ can be interpreted as the projection of the net angular momentum of the exchanged virtual photons responsible for dipole-dipole interactions (i.e. $m_r = \pm(q - q')$). The deficit $q'$ is converted to mechanical rotation of the molecule, consistent with overall energy conservation, Eq. (11).

In addition to degenerate couplings resulting in changes in internal states as described above, for atoms in excited vibration modes of a common spherical well there are further degeneracies and couplings allowed by the energy conservation law Eq.(11b) and selection rule Eq. (12). For instance, the product state of two atoms, each with one quanta of vibration along $z$, can couple to the seven dimensional degenerate subspace of two quanta shared between the two atoms,

$$|n_1 l_1 m_1\rangle \quad |n_2 l_2 m_2\rangle = \{|0\ 1\ 0\rangle \quad |010\rangle, |01-1\rangle \quad |011\rangle, |011\rangle \quad |01-1\rangle,$$
$$|020\rangle \quad |000\rangle, |000\rangle \quad |020\rangle, |100\rangle \quad |000\rangle, |000\rangle \quad |100\rangle\}. \tag{13}$$

All of these features must be accounted for if we are to utilize the dipole-dipole interaction for coherent quantum state manipulation and logic gates, as we consider in the next section.

## III. QUANTUM LOGIC GATES

### A. General considerations.

The dipole-dipole interaction discussed in Sec. II can be employed to bring about coherent interactions between neutral atoms. A useful paradigm to generate an arbitrary unitary evolution of a many-body system is the "quantum circuit", in which a series of logic gates act on a set of two-level quantum systems (qubits) [4], here the trapped atoms with pure states identified as the logical-$|0\rangle$ and $|1\rangle$. A crucial component is the two-qubit logic gate whereby the state of one atom (called the target) is evolved conditional on the logical state of the other (called the control). A familiar example is the "controlled-not" (CNOT), whereby the logical state of the target is flipped, $|0\rangle \quad |1\rangle$, *iff* the control qubit is in the logical-$|1\rangle$, and no change is made otherwise. Other examples that have no classical analog include the "controlled-phase" (CPHASE) and "square-root of swap" ($\sqrt{\text{SWAP}}$) gates [11]. In the former, the two-qubit state with both atoms in the logical-$|1\rangle$ acquires a *probability amplitude* of $-1$, $|1\rangle \quad |1\rangle \quad -|1\rangle \quad |1\rangle$, and nothing otherwise. The latter, if operated twice, swaps the state of the logical-$|1\rangle$ and logical-$|0\rangle$; operated once we have an equally weighted superposition of no-swap and swap. These examples are entangling two-qubit gates, which together with the ability to effect arbitrary single qubit transformations, can generate an arbitrary entangled state of the many-body system through sequential pairwise application. In the language of quantum information processing, these logic gates form a universal set for quantum computation [32].

We consider the design of such logic gates in optical lattices. A suitable three dimensional geometry is illustrated in Fig. 2. Confinement along each of the Cartesian axes is provided by a standing wave of far blue-detuned laser light. We take the laser frequency along each direction to be sufficiently different to eliminate interference between the standing waves. In the $x$-$y$ plane atoms are trapped in tubes at the nodes of the field arranged in a square array of periodicity $\lambda_T / 2$. Confinement along $z$ is provided by laser beams that are linearly polarized in the $x$-$y$ plane with a relative angle $\theta$ between the polarization vectors. These beams produce a pair of $\sigma_+ - \sigma_-$ standing waves with antinodes separated by $z = \lambda \tan^{-1}\left[\left(\tan \theta\right) / 2\right]$. The resulting optical potential from this lin-$\theta$-lin polarization configuration [33] permits us to trap two distinguishable sets of atoms: those most light-shifted by $\sigma_+$ light, and those who are most light-shifted by $\sigma_-$ light. We refer to these as the "($\pm$)-species". Near the minima, the optical potentials are approximately harmonic, with the curvature along each dimension determined by the beam intensities and detunings and by the oscillator strengths of the particular transitions driven by the fields.

The ability to dynamically control the angle $\theta$ between polarizations, and thus vary the distance between the nodes of the interleaved standing waves, allows one to separate and move atoms of the ($\pm$)-species relative to each other. Two atoms of different species, initially separated along the $z$-axis by $n \lambda_L / 4$ when the polarization configuration is lin$\perp$lin, can be made to overlap by rotating $\theta$ by $n \pi / 2$. If this angular rotation is made adiabatic with respect to the oscillation frequency as discussed below, then the center of mass wave functions of the atoms are unchanged. Once the atoms have been brought together they can be made to interact by applying an auxiliary laser pulse, referred to as the "catalysis field", which excites the atomic dipoles for a time necessary to achieve the desired two-qubit logical operations. Afterwards, the atoms can be separated by further adiabatic rotation of the lattice polarizations so that they no longer interact. The catalysis field is assumed to be tuned closer to resonance than the lattice fields and induces stronger dipoles (though still with saturation $s \ll 1$), so that the dipoles induced by the lattice fields can be neglected in their contribution to the atom-atom interactions.

To implement quantum logic in this scheme we define separate computational basis sets for the ($\pm$)-species. For example, one can choose

$$
\begin{aligned}
|1\rangle_\pm &= |F_\mp, M_F = \pm 1\rangle \quad | \quad _\pm\rangle_{ext}, \\
|0\rangle_\pm &= |F_\mp, M_F = \mp 1\rangle \quad | \quad _\pm\rangle_{ext},
\end{aligned}
\tag{14}
$$

where $F_\mp = I \pm 1/2$ are the two hyperfine levels associated with the $S_{1/2}$ ground state, $M_F$ is the magnetic sublevel, and $| \quad _\pm\rangle_{ext}$ is the external coordinate wave function of the associated potential. Under the condition that the lattice fields are very far detuned compared to the hyperfine splittings of the D2 resonance $|S_{1/2}\rangle \quad |P_{3/2}\rangle$ (we assume the detuning is small compared to the much larger fine structure splitting), the optical trapping potential takes the form of an effective magnetic field, with no coherences $M_F = \pm 2$ [21]. The magnetic quantum numbers $M_F$ are thus good quantum numbers and are preserved during rotation of the lattice polarizations. The large detuning limit of the trapping fields also makes the basis of Eq. (14) relatively robust to fluctuations in the lattice potentials and ambient magnetic fields. Since the Landé g-factors for the $|F_\mp\rangle$ and $|F_\mp\rangle$ are equal and opposite in sign (to within $\sim 10^{-3}$), the logical-$|1\rangle$ and logical-$|0\rangle$ experience nearly identical effective Zeeman shifts. Thus, fluctuations in the true magnetic fields and/or the effective magnetic field associated with the optical lattice potential will cause minimal changes in the energy *difference* between these states.

Such fluctuations would otherwise rapidly lead to single-qubit phase errors. Furthermore, because we define separate basis states for the (+) and (–) species, all entanglement resulting from the quantum logic occurs between the spins of *different* atoms, and there is no entanglement within a single atom between its internal and external degrees of freedom. This is in contrast to the proposal of Jaksch *et al.* [15], whereby the logical-$|1\rangle$ and logical-$|0\rangle$ of a given atom follow different potentials, leading to the formation of nonlocal entangled "Schrödinger-cat" states of individual atoms. Such states could potentially be highly susceptible to phase errors caused by spatial inhomogeneities in the trapping potential.

For a given basis we can now define protocols for different quantum logic operations. Single bit manipulations (e .g. rotations of the qubit state on the Bloch sphere and read-out of the qubit state) can be implemented through conventional spectroscopy, such as coherent Raman pulses and fluorescence spectroscopy as demonstrated in ion traps [20]. Two qubit entanglement is achieved by inducing a conditional dipole-dipole interaction. We will consider two kinds of gates in Sec. III.B. A CPHASE gate can be implemented with the following protocol. A pair of atoms of opposite species are brought into close proximity through rotation of the lattice polarizations as described above. If the catalysis field is tuned near the $\left|S_{1/2}, F\right\rangle \rightarrow \left|P_{3/2}\right\rangle$ resonance with detuning small compared to the ground-state hyperfine splitting, then dipoles are induced *only* for atoms in the logical-$|1\rangle$ states. If there are no off-diagonal matrix elements of the dipole-dipole interaction in the chosen logical basis, and assuming the gate is performed on a time scale much faster than the photon scattering rate, this causes only a non-zero level shift of the logical basis state $|1\rangle_+ |1\rangle_-$, and zero level shift of all other logical basis states (see Fig. 3a). If the atoms are then allowed to freely evolve for a time $= \hbar \pi / \langle V_{dd} \rangle$, we have $|1\rangle_+ |1\rangle_- \rightarrow -|1\rangle_+ |1\rangle_-$ with no change to the other logical basis states, as required for a CPHASE. A similar protocol can be constructed to implement a $\sqrt{\text{SWAP}}$ gate. Through an appropriate choice of logical basis and catalysis field, one can induce dipole-dipole couplings which are only off-diagonal in the logical basis, of the type $|1\rangle_+ |0\rangle_- \quad |0\rangle_+ |1\rangle_-$. Applying the interaction for a time such that a $/2$ rotation occurs in this subspace, we obtain the desired gate (see Fig. 3c). For a general initial state of the two qubits, the result of the logic gate will be an entangled state.

In all of the protocols, it is necessary to devise interactions that minimize loss of fidelity due to photon scattering and coherent coupling outside the logical basis ("leakage"). One leakage channel arising from the dipole-dipole potential discussed in Sec. II originates from the coherent coupling between degenerate internal states. Generally, neither $M_{F1}, M_{F2}$ nor $M_F = M_{F1} + M_{F2}$ is conserved in this process. From Eq. (8) we have the selection rule, $M_F = \pm |q - q|$, where $q$ is the net projection of angular momentum of the exchanged virtual photon along the atomic quantization axis. As summarized by selection rule Eq. (12), any deficit in the angular momentum of the exchanged photons must be balanced by an excitation of mechanical rotation of the two-atom molecule. If such transitions are allowed by energy conservation, they can be suppressed through judicious choice of lattice geometry. Choosing the two-atom confinement to have azimuthal symmetry with respect to the quantization axis, a partial wave expansion of the relative coordinate probability distribution contains only terms $Y_2^0 ( , )$, ensuring that only the terms with $q = q$ survive in Eq. (7), and thus $M_F = 0$. Leakage channels can also be suppressed by breaking the symmetry that leads to the degeneracy. A sufficient magnetic field can define the quantization axis, providing a linear Zeeman splitting of the ground state magnetic atomic sublevels greater than the cooperative linewidth of these states. Processes that do not conserve the total $M_F$ are thus detuned out of resonance (Fig. 1b). Preservation of the *individual* quantum numbers $M_{F1}, M_{F2}$ requires a nonlinear Zeeman shift or ac-Stark shift in the ground state manifold to break the degeneracy (Fig. 1c).

One final leakage channel we address is coherent coupling into the excited state manifold. We must ensure that all population returns to the ground states after the logic gate is completed. One means to achieve this is to adiabatically connect the ground-manifold to the field-dressed levels. Consider the simplified basis of two 2-level atoms considered in Appendix A, for atoms at a fixed relative position and coupled by a dipole-dipole interaction of strength $V_c$. Adiabatic evolution requires the level splitting between the two-atom ground and first excited eigenstates, $\hbar \Delta - V_c$, to be sufficiently large compared to off-diagonal coupling caused by the changing catalysis excitation. Furthermore, as shown in Appendix A, if $|\Delta| >> |V_c|$, then the single atom light shift due to the catalysis field and the dipole-dipole potential are separable, resulting in the effective interaction Hamiltonian,

$$H_{\text{int}} = \frac{s}{2}\hbar\Delta \left( |1\rangle_+\langle 1| + |1\rangle_-\langle 1| \right) + V_{dd},$$
(15)

(we have assumed as before that the ground state hyperfine splitting is sufficiently large to allow us to neglect off-resonance interaction with the catalysis field). Since the light-shift interaction is separable for the two atoms, its effect can be compensated by single bit gates acting on each atom independently.

An alternative scheme for ensuring that the atoms return to their ground state is to work in the opposite limit, and apply sudden pulses, fast compared to $V_{dd}/\hbar$. As an example, consider the use of *maximally* excited dipoles to implement a CPHASE in manner similar to Ramsey interferometry (see Fig. 4). In this protocol atoms are brought together and a short $\pi$-pulse acts only on the $(+)$-species, bringing it to the excited state *iff* it is in the logical-$|1\rangle$ ground state, and leaving the $(-)$-species unaffected. The atoms then evolve freely. If an excitation exists in the $(+)$-atom (i.e. it started in the logical-$|1\rangle$ state) it will be coherently exchanged with the $(-)$-atom *iff* that atom also began in the logical-$|1\rangle$ ground state; otherwise the dipole-dipole interaction is not resonant and there is no exchange interaction. For an interaction time $\tau = \hbar\pi/\langle V_{dd}\rangle$ the (+)-species returns to the excited state with probability amplitude $-1$. A $(-\pi)$ pulse is then applied to the (+)-species, returning the atoms to the electronic ground states but with an accumulated minus sign for $|1\rangle_+|1\rangle_-$. To implement this scheme it is necessary to independently excite a transition of one species. For instance, for an atom with nuclear spin $I$=3/2 one could choose the logical basis

$$
\begin{aligned}
|1\rangle_{\pm} &= \left|F=1, M_F = \mp 1\right\rangle \otimes |\psi_{ext}\rangle, \\
|0\rangle_{\pm} &= \left|F=2, M_F = \pm 1\right\rangle \otimes |\psi_{ext}\rangle,
\end{aligned}
\tag{16}
$$

and address the different species by the catalysis polarization (see Fig. 4c). By applying a $\sigma_+$-polarized $\pi$-pulse, one excites only the (+)-species on the transition $\left|F=1, M_F = -1\right\rangle \rightarrow \left|F=0, M_F = 0\right\rangle$. An equivalent protocol in the form of a CNOT has been proposed by Lukin and Hemmer [34] for dipole-dipole interacting dopants in a solid state host. There, distinct atomic species with different resonant energies are considered for addressing only control or target atoms.

## B. Specific Implementations

In the following we consider three examples which demonstrate the flexibility available for designing quantum logic gates. We will assign a logical basis set such as that given in Eq. (14). When the atoms are excited by $\pi$-polarized catalysis light, the figure of merit is given by Eq. (4), with $q$=0. To complete our quantum logic protocol, we must choose the external coordinate wave function for our qubits. Considerations include maximizing the dipole-dipole figure of merit and minimizing coherent leakage due to degeneracies.

Let us first consider the case of two atoms in the vibrational ground state sharing a common well. Though spherical wells maximize the radial overlap for atoms in their ground state, the dominant term in the interaction tensor is $f_{00} = n_2(k_L r) Y_2^0(\ ) \sim Y_2^0(\ )/(k_L r)^3$, which is orthogonal to the isotropic relative coordinate Gaussian wave function. This multipole component is nonzero, however, for nonspherical geometries and for higher motional states of the atoms in spherical wells.

One suitable design is to use ellipsoidal wells. Consider an axially symmetric harmonic potential with two atoms in the vibrational ground state, each described by a Gaussian wave packet with widths $x = y = x_0$ and $z = z_0$. The figure of merit can be calculated numerically including radiation terms, as a function of $= k_L x_0$ and $_\parallel = k_L z_0$ as presented in [14]. In order to optimize the figure of merit, we consider an approximate analytic expression for for tight localization. The external coordinate wave function separates into center of mass and relative coordinates, with rms widths of the latter given by $_{x,rel} = \sqrt{2} x_0$ and $_{z,rel} = \sqrt{2} z_0$. Taking only the near field contribution to $H_{dd}$, where $\langle g_{00} \rangle_{ext}$ 1 and $\langle f_{00} \rangle_{ext}$ $-3 \langle P_2(\cos\ )/(kr)^3 \rangle_{ext}$, we have from Eq. (4)

$$\frac{1}{4} \langle f_{00}(r,\ ) \rangle_{ext} = -\frac{3}{4} \int d^3x | \ _{rel}(r,\ )|^2 \frac{P_2(\cos\ )}{(k_L r)^3}$$

$$= -\frac{3}{16\sqrt{\ }^2 \ _\parallel} \int_0 \frac{dr}{r} \int_{-1}^1 d(\cos\ ) P_2(\cos\ ) \exp\left[ -\left(\frac{\sin^2\ }{4x_0^2} - \frac{\cos^2\ }{4z_0^2}\right) r^2 \right].$$

$$\tag{17}$$

Though this integral appears to have a logarithmic divergence at $r = 0$, this is not the case since the angular integration goes like $r^2$ for small $r$ [35]. We can evaluate this expression by making the radial integral converge through the substitution $1/r \to 1/r^{(1+\epsilon)}$. After taking the limit $\epsilon \to 0$ we find

$$\kappa = \frac{1}{16\sqrt{\pi}\,\gamma^2\,\sigma_{\parallel}}\left[-2 - 3\frac{\sigma^2}{\gamma^2} + 3\frac{\sigma^3}{\gamma^3} + \pi\frac{\sigma^3}{\gamma^3}\tan^{-1}\frac{\gamma}{\sigma}\right], \qquad (18)$$

where $\sigma^{-2} = \sigma_{\parallel}^{-2} - \sigma_{\perp}^{-2}$. Keeping $\sigma_{\perp}$ fixed while maximizing with respect to the ratio $\sigma_{\parallel}/\sigma_{\perp}$ gives $\kappa_{max} \approx -8.5 \times 10^{-3}/\sigma_{\perp}^3$ for a ratio $(\sigma_{\parallel}/\sigma_{\perp})_{max} \approx 2.18$. The relatively small prefactor $\kappa$ can be attributed to two sources:  the rms width of the relative coordinate Gaussian wave function in three dimension is at least $\sqrt{6}$ times the rms for a single particle in 1D, and the overlap of the angular distribution of the dipoles and $P_2(\cos\theta)$ is imperfect. As an example, given tight localizations $z_0 = \lambda/60$, $x_0 = \lambda/136$, corresponding to Lamb-Dicke parameters $\eta_{\parallel} = 0.1$, $\eta_{\perp} = 0.05$, we have $\kappa \approx -68$.

A disadvantage of using two atoms in a common prolate ellipsoidal well is that the interaction potentials for different orientations of the relative coordinate destructively interfere with each other.  For instance, for parallel dipoles aligned along z, $V_{dd} \sim -2d^2/r^3$ when the internuclear axis is along the polarization, and $V_{dd} \sim d^2/r^3$ for perpendicular separations.  A possible solution is to use non-overlapping spherical wells, separated along $z$.  We know that as this separation goes to zero, the dipole-dipole interaction goes to zero.  We also know that $V_{dd} \sim 1/(kr)^3$ goes to zero as the separation goes to infinity.  Thus at some intermediary value of atomic separation, the interaction must be maximum.  For the case of two spherical wells separated by $\Delta z$, we can write the two particle external wave function as a product of isotropic ground state single particle Gaussians.  The wave function is separable in center of mass, $\mathbf{R} = (\mathbf{r}_1 + \mathbf{r}_2)/2$, and relative coordinate $\mathbf{r} = \mathbf{r}_1 - \mathbf{r}_2$,

$$\Psi(\mathbf{R},\mathbf{r}) = \frac{1}{\sqrt{\pi} R_c^{1/2}} e^{\frac{|\mathbf{R}|^2}{2R_c^2}} \; \frac{1}{\sqrt{\pi} r_c^{1/2}} e^{-\frac{|\mathbf{r} - z\hat{z}|^2}{2r_c^2}} \; , \tag{19}$$

with characteristic lengths, $R_c = \sqrt{\hbar/2m}$, $r_c = \sqrt{2\hbar/m}$. Since the relative coordinate wave function is azimuthally symmetric, it is valid to consider only the $Y_2^0$ piece of the coupling tensor as discussed in Sec III.A. The figure of merit follows as in Eq. (17),

$$\Lambda = \frac{e^{-(\bar{z}/2)^2}}{\sqrt{\pi}} \left( \frac{1}{8} + \frac{3}{4\,\bar{z}^2} \right) - \frac{3\,\mathrm{erf}\left(\bar{z}/2\right)}{4\,\bar{z}^3} \frac{1}{\eta^3} \, , \tag{20}$$

where $\bar{z} = z/x_0$, and $\eta = k_L x_0$ with $x_0$ the single particle 1D localization. The form of Eq. (20) can be verified in two limits. For $z \gg x_0$, $\Lambda \to -0.75(k_L z)^{-3}$, the expected figure of merit for two point dipoles separated by distance $z$, with dipole vectors aligned with the relative coordinate vector. For $z \ll x_0$, we find $\Lambda \to -(\eta \bar{z})^2/(80\sqrt{\pi}\,\eta^3)$, vanishing quadratically as the separation between wells goes to zero. A plot of $\Lambda$ is shown in Fig. 5. The figure of merit is maximized at $z_{max}/x_0 \approx 2.5$ where $\Lambda_{max} \approx -0.015/\eta^3$. For example, at $\eta = 0.05$, $\Lambda_{max} \approx -123$. This is almost twice as good as the scheme using overlapping ellipsoidal wells with the same minimum localization.

Separated wells also have the advantage of reducing unwanted elastic and inelastic scattering processes which are significant for atomic separations on the order of a few Bohr radii. Given wave function Eq. (19), the probability for two atoms to be separated by $r < a$ is

$$P(a) = \frac{1}{\bar{z}\sqrt{\pi}} \left( e^{-(\bar{z}+\bar{a})^2/4} - e^{-(\bar{z}-\bar{a})^2/4} \right) - \frac{1}{2}\left( \mathrm{erf}\left(\frac{\bar{z}-\bar{a}}{2}\right) - \mathrm{erf}\left(\frac{\bar{z}+\bar{a}}{2}\right) \right) \, , \tag{21}$$

where $\bar{a} \equiv a/x_0$. In the limit $\bar{a} \to \infty$, $P \to 1$, as required. Proper design of the logic gate will require that within the radius of inelastic processes, this probability is sufficiently small. We will elaborate on this and other collisional issues in Sec. IV.

Higher vibrational states of overlapping spherical wells can also be used to encode the qubit for controlled logic. For instance one quanta of vibration along $z$ in each atom could be considered to code for the logical-$|1\rangle$. This is ill suited as a logical basis, however, because of the problem of coherent leakage. The couplings given by selection rules Eq. (11) and (12) connect the logical basis to a seven dimensional degenerate subspace of two vibrational quanta shared between the atoms as described in Sec. II, Eq. (13). Many of these couplings can be avoided, however, if instead we choose the so-called stretched states of vibration. Consider the logical basis

$$
\begin{aligned}
|1\rangle_{\pm} &= \left| F, M_F = \pm 1 \right\rangle \left| n = 0, l = 1, m = 1 \right\rangle, \\
|0\rangle_{\pm} &= \left| F, M_F = \pm 1 \right\rangle \left| n = 0, l = 0, m = 0 \right\rangle.
\end{aligned}
\tag{22}
$$

The logical-$|1\rangle$ states are circularly oscillating vibrational states which are a superposition of transverse oscillations,

$$
\left| n = 0, l = 1, m = 1 \right\rangle = -\frac{1}{\sqrt{2}} \left( \left| n_x = 1, n_y = 0, n_z = 0 \right\rangle + i \left| n_x = 0, n_y = 1, n_z = 0 \right\rangle \right).
\tag{23}
$$

This state can be created by using two sets of Raman pulses at right angles to each other. With the atoms occupying the new logical basis, the $\sigma$-polarized catalysis field is applied to the transition $\left| S_{1/2} F \right\rangle \to \left| P_{3/2} F \right\rangle$ with detuning large compared to the oscillation frequency. This induces nearly equal dipoles for atoms in logical-$|1\rangle$ and $|0\rangle$ states (see Fig. 3b).

Matrix elements of the dipole-dipole operator can then be calculated using Eq. (10). Unlike the previous cases discussed, the interaction operator is not diagonal in the computational basis set, $\left\{ |0\rangle_+ \quad |0\rangle_-, |0\rangle_+ \quad |1\rangle_-, |1\rangle_+ \quad |0\rangle_-, |1\rangle_+ \quad |1\rangle_- \right\}$, but instead has the form

$$V_{dd} = \frac{7}{4}\hbar \begin{array}{cccc} 0 & 0 & 0 & 0 \\ 0 & 1 & -1 & 0 \\ 0 & -1 & 1 & 0 \\ 0 & 0 & 0 & 4/7 \end{array}, \qquad = \frac{-\left( {}_{FF}\, c^{F,1,F}_{1,0,1} \right)^4}{35\sqrt{\phantom{3}}^3}s \quad . \tag{24}$$

In addition, as dictated by the selection rules of Sec. II, the dipole-dipole interaction couples the logical basis states to a subspace of states with two shared quanta, $|n_1 m_1 l_1\rangle \quad |n_2 m_2 l_2\rangle = \{|011\rangle \quad |011\rangle, |022\rangle \quad |000\rangle, |000\rangle \quad |022\rangle\}$. The matrix elements are

$$\langle 022| \quad \langle 000|V_{dd}|011\rangle \quad |011\rangle = \langle 000| \quad \langle 022|V_{dd}|011\rangle \quad |011\rangle = -\hbar \quad /\sqrt{2}$$
$$\langle 022| \quad \langle 000|V_{dd}|022\rangle \quad |000\rangle = \langle 000| \quad \langle 022|V_{dd}|000\rangle \quad |022\rangle = 9\hbar \quad /4 \tag{25}$$
$$\langle 022| \quad \langle 000|V_{dd}|000\rangle \quad |022\rangle = -5\hbar \quad /4 \quad .$$

The couplings within the degenerate vibrational subspace of Eq. (25) describe an effective two-level system with coupling between the state $|011\rangle \quad |011\rangle$ and the symmetric state $\left(|022\rangle \quad |000\rangle + |000\rangle \quad |022\rangle\right)/\sqrt{2}$. The antisymmetric state is uncoupled and "dark" to the interaction. The effective Rabi frequency within the coupled subspace is exactly , thus there is a recurance time $=$ / for population in the vibrational state $|011\rangle \quad |011\rangle$. For this interaction time the unitary operator in the logical basis is

$$U = \exp\left(-iV_{dd} \quad /\hbar\right) = \begin{array}{cccc} 1 & & 0 & 0 & 0 \\ 0 & e^{i\ /4} & 1 & -i & 0 \\ 0 & \dfrac{}{\sqrt{2}} & -i & 1 & 0 \\ 0 & & 0 & 0 & 1 \end{array} \tag{26}$$

This is the $\sqrt{\text{SWAP}}$ gate universal to quantum logic [11]. The figure of merit for this gate is

$$-\frac{1}{4}\{1_+1_-|f_{00}|1_+1_-\rangle = \frac{1}{140\sqrt{\eta}^3} \approx \frac{4.02\times10^{-3}}{\eta^3}, \qquad (27)$$

which for $\eta = 0.05$ gives $\approx 32$. This figure could be improved by a factor of 3.5 if atoms oscillating along $z$ were used instead. However, an anisotropy would have to be introduced into the trapping potentials to suppress couplings to degenerate states outside the logical basis. The flexibility of the optical lattice for performing different universal logic gates such as the CPHASE and $\sqrt{\text{SWAP}}$ may be useful in the implementation of complex instruction sets that optimize computations for algorithms using multiple entangling gates [36].

### C. Gate Fidelity

The value of the figure of merit $\kappa$ implies an absolute upper limit on the fidelity $F$ of entangling operations performed with a given quantum gate. This limit can be regarded as "fundamental", in the sense that it derives from a decoherence mechanism intrinsic to our scheme. In the following we examine how the upper limit on $F$ scales with lattice parameters and estimate its value in a first generation experiment. Ignoring for the moment the possibility of inelastic atomic collisions, the gate fidelity is limited by spontaneous light scattering not just from the catalysis field, but also from the lattice field. The probability of scattering a catalysis photon during one operation of the gate can in principle be suppressed to an arbitrary degree (large $|\eta|$) by tight localization of the wave packets. Tighter localization, however, requires a deeper lattice. Because there will always be a finite amount of laser power available to form the lattice this can be accomplished only at the cost of decreased lattice detuning, which in turn increases the probability of scattering a lattice photon during the gate operation. There exists then an optimum choice of lattice detuning, where the overall probability of error due to spontaneous light scattering is minimized.

We take the duration of a gate operation to be the time required to bring a pair of atoms together, perform the entangling operation, and return the atoms to their original positions. The catalysis field is present only during the entangling operation, of duration $\tau = \hbar/|\langle V_{dd}\rangle|$. It then follows from the definition of the figure of merit, $\kappa = \langle V_{dd}\rangle/\langle\gamma_{tot}\rangle$, that the probability of *not* scattering a catalysis photon is $\mathbf{F}_C = \exp(-\gamma_{tot}\tau) = \exp(-1/|\kappa|)$. The lattice can induce spontaneous photon scattering at any time throughout the duration of the gate, $T = t + \tau$, where $t$ is the time needed to transport the atoms together and apart again. The probability of *not* scattering a lattice photon is then $\mathbf{F}_L = \exp[-\gamma_L T]$, with $\gamma_L$ the rate of scattering from the lattice. Because the lattice and catalysis detunings are very different we can treat scattering from the two fields as independent processes. The overall gate fidelity is then the product

$$\mathbf{F} = \mathbf{F}_C\mathbf{F}_L = \exp(-1/|\kappa|)\exp(-\gamma_L T). \tag{28}$$

A lower bound on the time $\tau = \hbar/|\langle V_{dd}\rangle|$ needed to carry out an entangling operation is set by the use of maximally excited dipoles as described in Sec. III.A. Noting that the superradiant decay rate is at most $2\gamma$ and the dipole-dipole interaction is at most $\langle V_{dd}\rangle = 2\hbar\gamma$, we find

$\tau \geq 1/(2|\kappa|)$. A rough lower bound can be established also for the time $t$ needed to bring the atoms together and apart. One can move the atoms by adiabatic translation of the potentials, so as to avoid excitation of vibrational motion in the optical potential wells. This would require $t \gg 2\omega_{osc}^{-1}$. However, the authors in [15] have shown that for harmonic potential wells there exist optimized trajectories which allow the atoms to move faster than adiabatically, yet remain in the vibrational ground state at the initial, overlapped and final positions. With this scheme it may be possible to achieve $t = 4\,\omega_{osc}^{-1}$, independent of the original separation between the atoms. In either case the oscillation period in the lattice constitute a natural unit of time, and we write $t = n\,\omega_{osc}^{-1}$, where $n \gtrsim 2$ [37]. Even for the case of fast, non-adiabatic transport, the minimal duration $\tau \sim \hbar/(2|\kappa|)$ of the entangling operation contributes negligibly to the total gate time, which is roughly $T \approx n\,2\pi\,\omega_{osc}^{-1}$.

When the atomic qubits are brought together and the entangling operation performed, the lattice consists of three linearly polarized standing waves, and the oscillation frequency and localization in the optical potential is given by [38]

$$\frac{\hbar\,\omega_{osc}}{E_R} = \frac{\kappa^{-2}}{\sigma} = 4\sqrt{\frac{2}{3}\frac{U_1}{E_R}} , \tag{29}$$

where $U_1$ is the light shift from a single lattice beam, assuming unit oscillator strength (for simplicity we assume all six lattice beams have identical light shifts). The single-beam light shift is [38]

$$U_1 = s_1 \frac{\hbar\,\delta_L}{2} = \frac{\hbar}{8} \frac{I_1/I_0}{\delta_L/\Gamma} \tag{30}$$

where $s_1$ and $I_1$ are the single-beam saturation and intensity, $\delta_L$ is the lattice detuning, $I_0$ is the saturation parameter, defined so that $2\Omega^2/\Gamma^2 = I/I_0$. For $|\kappa| = c\,\sigma^{-3}$, where $c$ is a constant depending on the specific protocol, one finds

$$1 - \bar{F}_C \equiv \bar{\epsilon} = \frac{\Gamma}{8c} \left( \frac{12 E_R}{\hbar} \right)^{3/4} \left( \frac{I_1}{I_0} \right)^{-3/4} \left( \frac{\delta_L}{\Gamma} \right)^{3/4} . \tag{31}$$

We next estimate the average rate $\gamma_L$ of photon scattering induced by the lattice during a gate operation, for an atom in one of the logical basis states of Eq. (14). For a detuning $\Delta_L$ much larger than the excited state hyperfine splitting, the photon scattering rate $\gamma$ and light shift $U$ for a given hyperfine ground state are always related as $\hbar \gamma / U = \Gamma / \Delta_L$. The light shift is identical for the two logical states of the qubits, and this holds also for the scattering rate. In the three-color, blue detuned lattice of Fig. 2, an atom of e. g. the $(+)$–species is always trapped around the nodes of the standing waves in the $x$-$y$ plane, and around the nodes of the $\sigma_+$ standing wave component along the $z$-axis. Averaging over the finite extent of the atomic wave packet one finds that the scattering rate from these standing waves is suppressed by a factor $\eta^2 \ll 1$ relative to the maximum value at the antinodes. This suppression of photon scattering is the primary motivation to work in blue-detuned lattices. During a typical gate operation, however, the atom is moved by many wavelengths along the $z$-axis, and the average induced scattering rate from the $\sigma_-$ standing wave component will be close to half its maximum value. For tightly localized wavepackets this rate is much larger than the scattering rate from the transverse and $\sigma_+$ standing waves, and the latter can be ignored. The average scattering rate for a $(+)$–species atom during a gate operation is then

$$\gamma_L = \frac{U_1}{\hbar}\frac{\Gamma}{\Delta_L}\overline{F}\left(O_{FF}\, c^{F,1,F}_{M_F,-1,M_F-1}\right)^2 = \frac{U_1}{\hbar}\frac{\Gamma}{\Delta_L}\left(\frac{2}{3}-\frac{1}{3F}\right). \tag{32}$$

Combing Eqs. (29) and (32) it is straightforward to show that

$$1-\mathcal{F}\approx\gamma_L T = \frac{\sqrt{3}}{8}\left(\frac{2}{3}-\frac{1}{3F}\right)\sqrt{\frac{\hbar\omega}{E_R}}\,n\left(\frac{I_1}{I_0}\right)^{1/2}\left(\frac{\Gamma}{\Delta_L}\right)^{-3/2}. \tag{33}$$

Substituting Eqs. (31) and (33) in Eq. (28) and maximizing with respect to $\Delta_L$ yields a maximum fidelity or minimum error-probability of

$$1 - \mathcal{F} \approx \frac{n}{c^2} \left( \frac{2}{3} - \frac{1}{3F} \right) \left( \frac{E_R}{\hbar} \frac{I_0}{I_1} \right)^{1/3} \tag{34}$$

at a detuning of

$$\Delta_L = \frac{c^{4/9}}{(12)^{1/9}} \left( \frac{2}{3} - \frac{1}{3F} \right)^{4/9} \left( \frac{\hbar}{E_R} \right)^{5/9} n^{4/9} \left( \frac{I_1}{I_0} \right)^{5/9}. \tag{35}$$

For example, for cesium with $F = 4$, $\hbar \gamma / E_R = 2.5 \times 10^3$, excited with a large but not unrealistic intensity of $I_1 = 10^5 I_0$ ( 0.5 W of optical power in a beam with spotsize $w = 0.6$ mm), and for the protocol of optimally separated spherical wells, with $c = 0.015$ (Eq. (20)) and $n=2$, we estimate a fidelity $\mathcal{F} \approx 0.92$ at an optimal detuning $\Delta \approx 6 \times 10^3 \gamma$. This is sufficient to permit a non-trivial number of gate operations in initial experiments, and to produce entangled states with multiple atoms. Improvements in fidelity can be made with increased laser intensity, though from the scaling in Eq. (40) it is clear that very high-Q build-up cavities are required if substantial improvements are to be achieved. Ultimately, significant improvement of the fidelity will depend on improving the scaling parameter $c$ and localization $\sigma$, which is constrained here by the approximations made in our model (e.g. detuning independent figure of merit) and other aspects of our specific protocol, as we discuss in the next section.

# IV. OUTLOOK

We have explored the possibility of creating multiparticle entanglement using induced dipole-dipole interactions between pairs of alkali atoms in an optical lattice. This system offers tremendous flexibility for quantum control of both internal and external degrees of freedom. By designing interactions that can be selectively turned on and off through an external control field, coherent interactions between atoms can be induced with high fidelity, opening possibilities for designing quantum logic gates to perform quantum information processing tasks.

The fidelity of the two-qubit logic gate, finite due to spontaneous emission, was estimated in Sec. III with modest CW laser power, and is sufficient to allow us to create entangled-states of multiple atoms. This holds great promise for a variety of applications in precision measurement [1] and quantum simulations [3]. It is important to note that these applications are possible *in the short term* given ensemble preparation and measurement on sparse, randomly filled lattices as are available in today's laboratory experiments.

In the much longer term, the promise of universal fault-tolerant quantum computing places very strong constraints on the physical system. Some of the requirements are technical challenges for our protocol. For example, we must address individual atomic qubits, necessary both during unitary evolution and in the final readout, and achieve controlled filling of the optical lattice. We can, however, perform detailed studies of our two-qubit logic gate in the short term entirely through ensemble measurements, which would establish an important proof of principle and act as a guide towards improved gate design. A discussion of how to measure the fidelity of a quantum gate with ensemble measurements in a sparsely filled lattice is given in Appendix C. Other requirements are more fundamental. Specifically, fault-tolerant quantum computation demands extremely low error rates, e.g. $1 - \mathcal{F} < 10^{-4}$ [39] or $10^{-3}$ for some models [40]. It is doubtful that our proposed protocols can reach these thresholds, even with the best high-Q optical power build-up cavities. Fortunately we see possibilities for substantial improvements by extending the theoretical analysis beyond the simplifying assumptions considered up to this point.

Given the scaling arguments of Sec. I, the dipole-dipole figure of merit has the form $\sim c / ^{3}$. One may consider the ultimate limits on these parameters given practical considerations. In our simple protocol, the localization is limited by lattice intensity and detuning, such that spontaneous photon scattering from the trapping lattice is no greater than from the catalysis field. This places severe constraints on the maximum fidelity since a good fraction of the error probability occurs during periods when the atoms do not interact. A possible solution would be to maintain the atoms in relatively shallow lattices, *except* during two-qubit operations. For example, if the lattice intensity is increased suddenly from $I_i$ to $I_f$, then in the harmonic approximation the atom wavepackets will "squeeze" from an initial localization of $_0$ to a minimum uncertainty of $_{min} = {}_0 \left( I_i / I_f \right)^{1/2}$ [33]. This should be contrasted with an adiabatic compression of the atomic wavepacket for which $_{min} = {}_0 \left( I_i / I_f \right)^{1/}$. "Turning on" localization on demand has the potential for dual benefits: the probability of scattering from the trapping lattice is greatly reduced and localization can be substantially increased for the same laser power and detuning. An analysis along the lines presented in Sec. III.C indicates that such a protocol has the potential to reduce the error rate by one or two orders of magnitude. To make use of the tight localization, the dipole-dipole interaction must be shorter than the "breathing period" of the squeezed wavepacket, or $|V_{dd}| >> 2 \, _{osc}$. This can be accomplished using a highly-exicted dipole protocol as described in Sec. III. In a practical experiment, one will be constrained by anharmonicity in the initial potential. This will limit the minimum localization and introduce new leakage channels related to the "recapture" of the squeezed atomic wavepacket in the initial vibrational state of the defined logical basis state. Methods in coherent control of atom wavepackets may provide the means to overcome this limitation [41].

The remaining parameter $c$ is more specific to the details of the protocol and the approximations of our analysis. Under the assumptions of our model, $c$ is determined solely by geometry and might also benefit from wavepacket engineering to maximize the relevant dipole-dipole multipole component of the relative coordinate probability distribution. More importantly, we must address the limitations of our model. For atoms separated by distances on the order of an optical wavelength it is appropriate to consider molecular rather than atomic resonances [23]. We have avoided explicit calculation of the molecular potentials by assuming a detuning that is large with respect to the splitting of these potentials. Though this assumption greatly simplifies the analysis, it may not be the optimal operating point for the system. Furthermore, at such a large detuning approaching the hyperfine splitting of the ground state, one cannot necessarily induce a dipole only for atoms in the logical-$|1\rangle$ states as we have assumed for some protocols.

Extending our model to include molecular resonance could have strong impact on the dipole-dipole figure of merit. For example, we have argued that for atoms in the vibrational ground state of a common isotropic spherical well, the dipole-dipole coherent interaction is zero due to destructive interference when integrating over all angles of the relative coordinate vector. However, at finite detuning, e.g. red of atomic resonance, the catalysis field will preferentially excite the attractive potential, leading to finite interaction, and an increase in the parameter $c$. Another example is the use of subradiant states. We have implicitly assumed that our dipoles are excited in phase, leading to Dicke superradiance. However, molecular resonances exist for dipoles oscillating out of phase, which might be excited with a sufficiently intense catalysis field. This too would impact the maximum possible value of $c$ in the figure of merit.

Finally, the assumption of large detuning is brought into question when considering two atoms occupying the same well, where for some small internuclear separations the catalysis field (and the lattice field for that matter) will be *resonant* with the molecular potentials (the so called Condon point). Inelastic processes are highly probable in this case. Many of these inelastic processes should be suppressed by utilizing highly localized, but separated atoms as were considered in Sec. III. Careful choice of the parameters can avoid Condon points over the extent of the relative coordinate probability density. In this regime off-resonant excitation dominates with an $r$ dependent detuning from the excited molecular eigenstates (see Fig. 6). Furthermore, the internuclear axis acquires a specific orientation with respect to the direction of the excited dipoles, and the different molecular potentials must be weighted accordingly. This is a regime not usually encountered in studies of photo-associative collisions. One future task is to include the molecular potentials in a full analysis. Such a model will present considerable challenges, especially when including the complex internal hyperfine structure of the alkalis [24].

Consideration of collision phenomena for atoms in tightly localized traps opens the door to a host of novel phenomena. Examples includes a breakdown of the scattering length approximation for electronic ground-state s-wave scattering [42] and the production of bound molecules though photoassociation, or a transition through a Feshbach resonance [43]. Recent experiments have been performed in Bose-Einstein condensates in which colliding pairs of atoms were resonantly transferred to an bound electronic ground state molecular resonance via Raman laser pulses [44]. This molecular bound state could act as an auxiliary level for performing a CPHASE gate by applying a $2\pi$ laser pulse between the free atom computational basis state and the molecular resonance. Whichever protocol ultimately holds the greatest promise, the rich structure of the neutral-atom / optical-lattice system provides new avenues for explorations of quantum control and information processing.

**Acknowledgements**

We gratefully acknowledge helpful discussions with Carlton Caves, Paul Alsing, John Grondalski and Shohin Ghose. We are especial thank Gary Herling for introducing us to the Moshinski-Bracket formalism and Aephraim Steinberg for pointing out the proper protocol for entangling atoms with maximally excited dipoles. IHD and GKB acknowledge support from the National Science Foundation (Grant No. PHY-9732456). PSJ was supported by the NSF (Grant No. PHY-9503259), by the Army Research Office (Grant No. DAAG559710165), and by the Joint Services Optics Program (Grant No. DAAG559710116).

**References**

[1] J. J. Bollinger et al., Phys. Rev A **54**, R4649 (1996). S. F. Huelga et al., Phys. Rev. Lett. **79**, 3865 (1997); A.M. Childs, J. Preskill, and J. Renes, (1999), Los Alamos preprint server quant-ph/9904021 v2 (1999).

[2] S. Lloyd, Science **273**, 1073 (1996).

[3] A. Sørensen and K. Mølmer, Phys. Rev. Lett. **83**, 2274 (1999).

[4] *Introduction to Quantum Computation and Information*, edited by H.-K. Lo, S. Popescu, and T. Spiller (World Scientific, Singapore, 1998).

[5] J. I. Cirac and P. Zoller, Phys. Rev. Lett. **74**, 4091 (1995).

[6] T. Pellizzari et al., Phys. Rev, Lett **75**, 3788 (1995). P. Domokos *et al.,* Phys. Rev. A **52**, 3554 (1995).

[7] C. Monroe *et al.,* Phys. Rev. Lett. **75**, 4714 (1995); Q.A. Turchette *et al.*, Phys. Rev Lett **81**, (1998).

[8] Q.A. Turchette *et al.*, Phys. Rev. Lett. **75**, 4710 (1995); E. Hagley *et al.*, Phys. Rev. Lett. **79** (1997).

[9]   N. A. Gershenfeld and I. L. Chuang, Science **275**, 350 (1997).  D. G. Cory *et al*., Proc. Natl. Acad. Sci. USA **94**, 1634 (1997).

[10] S.L. Braunstein *et. al.*,  Phys. Rev. Lett. **83**, 1054 (1999).

[11] D. Loss and D.P. DiVincenzo, Phys. Rev. A **57**, 120 (1998).

[12] J. E. Mooij *et al.*, Science **285**, 1036 (1999).

[13] B.E. Kane, Nature (London) **393**, 133 (1998);  R. Vrijen *et al.*, Los Alamos preprint server quant-ph/9905096 (1999).

[14] G.K. Brennen *et al*., Phys. Rev. Lett. **82**, 1060 (1999).

[15] D. Jaksch *et al.*, Phys. Rev. Lett. **82**, 1975 (1999).

[16] Q.A. Turchette *et al.*, Laser Physics **8**, 713 (1998).

[17] A. Hemmerich, Phys. Rev. A **60**, 943 (1999).

[18] S.E. Hamann *et al.*,  , Phys. Rev. Lett. **80**, 4149 (1998).

[19] D. Jaksch *et al.*, Phys. Rev. Lett. **81**, 3108 (1998).

[20] For a review of experiments that perform quantum coherent manipulations of trapped ions see, D. J. Wineland *et al.*, J. Res. Inst. Stand. Techol. **103**, 259 (1998).

[21] I. H. Deutsch and P. S. Jessen, Phys. Rev. A **57**, 1972 (1998).

[22] J. Schmeidmayer  Eur. Phys. J. D **4**, 57 (1998);  E. A. Hinds, M. G. Boshier, and I. G. Hughes, Phys. Rev. Lett. **80**, 645 (1998), M. Drndic et al., Appl. Phys. Lett. **72**, 2906 (1998).

[23] For a review see J. Weiner *et al*., Rev. Mod. Phys.  **71**, 1 (1999), and references therein.

[24] C. J. Williams and P. S. Julienne J. Chem. Phys. **101** 2634 (1994); P. D. Lett, P. S. Julienne, and W. D. Phillips, Annu. Rev. Phys. Chem. **46**, 423 (1996).

[25] R. H. Lemberg, Phys. Rev. A **2**, 883 (1970);  M. Trippenbach, B. Gao, J. Cooper, and K. Burnett, Phys. Rev. A **45**, 6555 (1992).

[26] E. U. Condon and G. H. Shortley, *The Theory of Atomic Spectra*, (Cambridge University Press, 1935).

[27] J. Guo and J. Cooper, Phys. Rev. A **51**, 3128 (1995); E.V. Goldstein, P. Pax, and P. Meystre, Phys. Rev. A **53**, 2604 (1996).

[28] P.S. Julienne, Phys. Rev. Lett. **61**, 698 (1988).

[29] C. Cohen-Tannoudji, B. Diu, F. Laloe, *Quantum Mechanics   vol. 1*, (Hermann, Paris, France, 1977).

[30] M. Moshinksy, Nuclear Physics **13**, 104 (1959).

[31] T. A. Brody, G. Jacob, and M. Moshinsky, Nuclear Physics **17**, 16 (1960).

[32] A. Barenko *et al.,* Phys. Rev. A **52**, 3457 (1995).

[33] S. Marksteiner *et al.*, Appl. Phys. B **B60**, 145 (1995).

[34] M. D. Lukin and P. R. Hemmer, Los Alamos preprint server quant-ph/9905025 (1999).

[35] M. G. Moore, P. Meystre, Phys. Rev. A **56**, 2989 (1997).

[36] G. D. Sanders, K. W. Kim, and W. C. Holton, Phys. Rev. A **59**, 1098 (1999); G. Burkard, D. Loss, D. P. DiVincenzo, and J. A. Smolin, Los Alamos preprint server quant-ph/9905230 (1999).

[37] As the atoms are transported in the lattice by translating the $\pm$ standing waves relative to each other, the oscillation frequency is modulated by a factor $\sqrt{2F}$, from the maximum value in Eq. (29) when the lattice configuration along $\hat{z}$ is lin||lin, to a minimum value $4\sqrt{U_1/3F E_R}$ when the configuration is lin⊥lin.  This must be compenstated for if atoms are to be transported non-adiabatically.

[38] P. S. Jessen and I. H. Deutsch, Adv. Atom. Mol. Opt. Phys. **37**, 95 (1996).

[39] J. Preskill, Proc. R. Soc. London, Ser. A **454**, 385 (1998).

[40] C. Zalka, Los Alamos preprint server quant-ph/9612028 (1996).

[41] See for example, *Mode Slective Chemistry*, edited by J. Jortner, R. D. Levine, and B. Pullman (Dordrecht, Boston, 1991).

[42] E. Tiesinga *et al.,* APS Centennial Meeting, Atlanta, Session SB13.09 (1999).

[43] Ph. Courteille *et al.,* Phys. Rev. Lett. **81**, 69 (1998).

[44] R.S. Freeland *et al.,* APS Centennial Meeting, Atlanta, Session OB16.05 (1999)

## APPENDIX A

Consider two 2-level atoms with ground and excited states $|g\rangle, |e\rangle$ interacting with a laser field and the vacuum.  After tracing over the vacuum modes, the effective non-Hermitian Hamiltonian is

$$H_{eff} = H_A + H_{AL} + H_{dd},$$ (A1)

where in the rotating wave approximation, the dressed atomic Hamiltonian is

$$H_A = -\hbar \left( -i\frac{\gamma}{2} \right) \left( |e_1\rangle\langle e_1| \quad \hat{\mathbf{1}}_2 + \hat{\mathbf{1}}_1 \quad |e_2\rangle\langle e_2| \right),$$ (A2)

and the atom-laser interaction is

$$H_{AL} = -\frac{\hbar}{2} \left[ \left( |e_1\rangle\langle g_1| + |g_1\rangle\langle e_1| \right) \quad \hat{\mathbf{1}}_2 + \hat{\mathbf{1}}_1 \quad \left( |e_2\rangle\langle g_2| + |g_2\rangle\langle e_2| \right) \right].$$ (A3)

From Eq. (1b), reduced to a two level system with $\mathbf{D}_i = \mathbf{e}_q |g_i\rangle\langle e_i|$, the dipole-dipole coupling interaction is

$$H_{dd} = V_c - i\frac{\hbar\gamma_c}{2}\left[|e_1 g_2\rangle\langle g_1 e_2| + |g_1 e_2\rangle\langle e_1 g_2|\right], \tag{A4}$$

where $V_c = -\hbar\langle f_{qq}\rangle/2$ and $\gamma_c = \langle g_{qq}\rangle$. Going to the "molecular-basis" of eigenstates of $H_0 = H_A + H_{dd}$,

$$|g_1 g_2\rangle, \quad |\pm\rangle = \frac{|e_1 g_2\rangle \pm |g_1 e_2\rangle}{\sqrt{2}}, \quad |e_1 e_2\rangle, \tag{A5}$$

the Hamiltonian is $H_{eff} = H_0 + H_{dd}$, with

$$H_0 = -\hbar(\gamma + i\delta/2)(2|e_1 e_2\rangle\langle e_1 e_2| + |+\rangle\langle+| + |-\rangle\langle-|) \tag{A6a}$$
$$+ (V_c - i\hbar\gamma_c/2)(|+\rangle\langle+| - |-\rangle\langle-|),$$
$$H_{AL} = \frac{\hbar\Omega}{\sqrt{2}}(|g_1 g_2\rangle\langle+| + |e_1 e_2\rangle\langle+| + h.c.). \tag{A6b}$$

The symmetric state $|+\rangle$ is superradiant with linewidth $\gamma + \gamma_c$, coupling to $|g_1 g_2\rangle$ and $|e_1 e_2\rangle$ with Rabi frequency $\sqrt{2}\Omega$. The state $|-\rangle$ is subradiant with linewidth $\gamma - \gamma_c$.

Treating $H_{AL}$ as a perturbation to $H_0$ we find the ground state energy to second order,

$$E_{gg} = \frac{\hbar\Omega^2/2}{(\delta - V_c/\hbar) + i(\gamma + \gamma_c)/2}. \tag{A7}$$

In the limit $\delta \gg V_c, \gamma$, the saturation is independent of the dipole-dipole interaction and

$$E_{gg} \approx s(\hbar\delta - i\hbar\gamma/2) + s(V_c - i\hbar\gamma_c/2), \tag{A8}$$

where $s = \Omega^2/(2\Delta^2)$ is the single atom saturation parameter. The first term in Eq. (A8) represents the sum of the single atom light-shifts and the photon scattering rate. The second term represents the dipole-dipole level-shift of the ground state and the contribution to the linewidth arising from cooperative effects. . These level shifts of the molecular eigenstates in this perturbative limit are shown in Fig. A1.

For an atom with a manifold of ground states, as considered in the text, the cooperative contribution to the eigenvalue is replaced by the effective Hamiltonian Eq. (5a). The figure of merit is defined as

$$Q = \frac{V_c}{-2\mathrm{Im}(E_{gg})} = \frac{-\langle f_{qq}\rangle}{2\left(1 + \langle g_{qq}\rangle\right)},\tag{A9}$$

depending only on geometry and independent of detuning in agreement with Eq. (4) in the text.

## APPENDIX B

In this appendix we review a method for calculating the tensor coupling between two particles (here two atoms) occupying a common spherical well, originally developed in applications in nuclear physics [30]. Each particle is in a vibrational eigenstate specified by quantum numbers in the spherical coordinates, $|n_i, l_i, m_i\rangle$. The product state of the two-body system is first expressed in terms of the coupled angular momentum representation, with total angular momentum and $z$-axis projection quantum numbers $\lambda, \mu$, as in Eq. (9). Given an interaction potential that depends only on the relative separation coordinate, the coupling is calculated by going to the basis of relative and center of mass states. Introducing the states $|nl, NL, \lambda\rangle$, where $n,l(N,L)$ are the radial and angular momentum quantum numbers of the relative(center of mass) coordinates of the two-body system, the expansion is

$$|n_1 l_1, n_2 l_2; \ \mu\rangle = \sum_{n \, l \, N \, L} \langle n \, l \, N \, L, \ |n_1 l_1, n_2 l_2, \ \rangle |nl, NL, \ \mu\rangle. \tag{B1}$$

The expansion coefficients $\langle n_1 l_1 \, n_2 l_2, \ |nl, NL, \ \rangle$ are known from nuclear physics as "Moshinsky brackets", which are tabulated real coefficients found using recursion relations [30], and are independent of the projection quantum number $\mu$. The brackets are non-zero if conservation of energy is satisfied,

$$2 n_1 + l_1 + 2n_2 + l_2 = 2 N + L + 2n + l. \tag{B2}$$

We thus seek an expression for the matrix element of the tensor coupling $T_K^Q(\mathbf{r}) \ \hat{\mathbf{1}}_{cm}$, where $\hat{\mathbf{1}}_{CM}$ is the identity operator acting on the center of mass degrees of freedom. For the case at hand the dipole-dipole coupling has the form $T_K^Q(\mathbf{r}) = V(r) Y_2^{m_r}( \ , \ )$. Using the Monshinsky expansion we have

$$\langle n_1 l_1, n_2 l_2, \ \mu | V(r) Y_2^{m_r}( \ , \ ) \ \hat{\mathbf{1}}_{CM} | n_1 l_1, n_2 l_2, \ \mu \rangle =$$
$$\sum_{\substack{n, l, N, L, \\ n', l', N', L'}} \langle n_1 l_1, n_2 l_2, \ |n \, l \, N L, \ \rangle \langle n' \, l', N' \, L', \ |n_1 l_1, n_2 l_2, \ \rangle$$
$$\langle nl, NL, \ \mu | V(r) Y_2^{m_r}( \ , \ ) \ \hat{\mathbf{1}}_{CM} | n' \, l', N' \, L', \ \mu \rangle \tag{B3}$$

These summations can be simplified. Since the potential couples only the relative coordinate states the center of mass quantum numbers are conserved, $N = N', L = L'$. Further, one free parameter is constrained by conservation of energy,

$$n' = n + n_1 + n_2 - n_1' - n_2' + \left( l_1 + l_2 - l_1' - l_2' + l - l' \right)/2. \tag{B4}$$

This reduces the interaction to a summation over five indices $n, l, N, L, l'$.

The remaining matrix element can be simplified using the $6$-$j$-symbol coupling rules and the Wigner-Eckart Theorem [25],

$$\langle nl, N L, \mu | V(r) Y_2^{m_r}(\ , ) \hat{\mathbf{1}}_{CM} | n' l', N L, \mu' \rangle =$$
$$(-1)^{L-\mu-l'} \sqrt{2l'+1}\, c_{\mu',m_r,\mu}^{l',2,l} \left\{ \begin{matrix} l & L \\ l' & 2 \end{matrix} \right\} \langle nl \| V(r)\hat{\mathbf{Y}}_2 \| n' l' \rangle, \tag{B5}$$

where the reduced matrix element involves only an integration over the radial coordinates,

$$\langle nl \| V(r)\hat{\mathbf{Y}}_2 \| n' l' \rangle = \sqrt{(2l'+1)5/(4\pi)}\, c_{0,0,0}^{l,2,l'} \langle n \| V(r) \| n' l' \rangle. \tag{B6}$$

The radial integral can be expressed as

$$\langle nl \| V(r) \| n' l' \rangle = \sum_{p=(l+l')/2}^{(l+l')/2+n+n'} B(nl, n'l', p) I_p(V(r)), \tag{B7}$$

where $B(nl, n'l', p)$ are radial function expansion coefficients given in [31], and $I_p(V(r))$ are the Talmi integrals,

$$I_p(V(r)) = 2/\Gamma(p+3/2) \int_0^\infty r'^{2p} e^{-r'^2} n_2(2\lambda_x r')r'^2 dr', \tag{B8}$$

where $\Gamma(x)$ is the gamma function, and $r' = r/\sqrt{2}$ is a dimensionless variable in units $\sqrt{\hbar/(m\omega)} = \sqrt{2}x_0$. Substituting Eqs. (B7) and (B6) into Eq. (B5), we obtain the expression given in Eq. (10) of the text.

# Appendix C

Characterization of a quantum logic gate requires that we measure the fidelity with which it achieves its truth table and determine the relative importance of various error mechanisms. In a sparse, randomly filled lattice, a reliable ensemble measurement of the truth table is complicated by the fact that only a small fraction of the qubits will be paired, i.e. have nearest neighbors on the lattice, compared to the much larger background of unpaired qubits. Nonetheless, it is possible to isolate the desired signal due to paired qubits from the background by first using the gate itself to identify the accidentally paired atoms, and then using radiation pressure to clear unpaired atoms from the lattice. Repeated application of the gate will then allow us to measure the error probability.

To emphasize this important point, we show how to perform an ensemble measurement of the truth table of a CPHASE gate. In practice it is much easier to detect changes in atomic populations than phase shifts, so the CPHASE gate is sandwiched between a pair of Hadamard single-bit rotations on the target qubit to achieve a CNOT gate in the usual manner [11]. For the purpose of illustration we will consider a specific input state $|1\rangle_+ \quad |0\rangle_-$ for the CNOT gate, where the $(+)$ and $(-)$ species act as control and target bits, respectively. Only qubits with a nearest neighbor are brought together and acted upon by the CNOT operation; we will assume the gate leaves the majority of unpaired qubits behind in their original state. For the paired qubits, the gate succeeds with probability $F < 1$, flipping the target bit to the $|1\rangle_+$ state while leaving the control bit unchanged. There are only three types of errors that can lead to sub-unit fidelity: (i) the control or target qubit ends up in the correct state, but its partner is lost outside the computational basis; (ii) both qubits are lost; (iii) the control and/or target qubit ends up in the wrong logic state. For a $|1\rangle_+ \quad |0\rangle_-$ input, correct operation of the gate leaves population in the $|1\rangle_+$    $|1\rangle_-$ state of the $F$ manifold, while errors populate the logical $|0\rangle_+$ and/or $|0\rangle_-$ states of the $F$ manifold, or states outside the computational basis.  We can "flush" qubits in the $F$ manifold with near unit efficiency from the lattice with radiation pressure from a laser beam tuned to the $F$      $F-1$ cycling transition, transfer population of the $F$ manifold outside the logical basis to the $F$ manifold by Raman pulses, and flush once more.  After we have flushed the failure modes of the gate in this fashion, all the failed gate operations have similar outcomes: one or both of the original, paired qubits are lost.  The flushing also serves to remove all qubits of the target species which were originally unpaired.

We can now determine the fidelity through a repeated measurement.  If we start out with $N$ pairs of qubits, the first CNOT/flushing cycle leaves $\mathcal{F}N$ qubits of the target species in pairs. In addition    $(1-\mathcal{F})N$ new unpaired qubits of the target species have been created by gate errors that affect the control, but not the target,      being the fraction of such errors. If we rotate the target qubits back to the logical-zero state and carry out the CNOT/flushing cycle once more, there will be $\mathcal{F}^2N$ paired and    $(1-\mathcal{F})\mathcal{F}N$ unpaired qubits of the target species.  The fidelity $\mathcal{F}$ is then simply the fractional decrease in the number of target species qubits between the first and the second CNOT/flushing cycle.  Similar sequences of CNOT/flushing cycles and single-qubit rotations allow to determine gate truth tables with other logical states as input.  Even if a small fraction of the unpaired qubits are flipped by the CNOT gate, it is still possible to extract the fidelity $\mathcal{F}$, though it may require one or more additional gate operations to eliminate a sufficient number of the original unpaired atoms.  It is also straightforward at any step to measure populations of hyperfine ground states outside the computational basis, in order to help distinguish between different failure mechanisms such as spontaneous emission and inelastic collision processes.

**FIGURE CAPTIONS**

**FIG. 1.** Fundamental photon exchange processes allowed in the dipole-dipole interaction. Solid lines indicate stimulated emission and absorption of a laser photon; wavy lines indicate emission and absorption of a virtual photon responsible for the exchange interaction. The contributing component of the dipole-dipole interaction tensor, $f_{qq}$, from Eq. (6) is indicated above. The black and white dots represent the initial state of the two interacting atoms; the gray and striped dots represent the final states respectively. (a) With degenerate grounds states, transitions which conserve neither $M_{F1}, M_{F2}$ nor total $M_F$ are allowed. (b) Under a linear Zeeman shift only processes conserving total $M_F$ are resonant. (c) A nonlinear Zeeman or ac-Stark shift will constrain the interaction to return both atoms to their initial states.

**FIG. 2.** Schematic of a 3D blue-detuned optical lattice. (a) Two pairs of -polarized beams provide transverse confinement, and the beams along $z$ in the lin- -lin configuration provide longitudinal confinement in $_+$ and $_-$ standing waves. (b) Potential surfaces for the ($\pm$)-atomic species, described in the text, shown here as in gray and white, are moved along the $z$-axis through a rotation of the angle between polarization vectors.

**FIG. 3.** Energy level structure of the logical basis associated with two-qubit logic gates. Basis states are denoted for ($\pm$)-species as described in the text. (a,b) CPHASE configuration: The "catalysis field" excites dipoles only in the logical- $|1\rangle$ states, chosen for both species in the upper ground hyperfine manifold $F$. The dipole-dipole interaction is diagonal in this basis and results solely in a level shift of the $|1\rangle_+$ $|1\rangle_-$ state. Operation of this gate with high fidelity requires this shift to be large compared to the cooperative linewidth. (c,d) $\sqrt{\text{SWAP}}$ configuration. The logical basis is encoded in the vibrational degree of freedom as described in Sec. III.B. For an appropriate choice of geometry and pulse timing, there is an off-diagonal coupling between the logical states $|1\rangle_+ \quad |0\rangle_-$ and $|0\rangle_+ \quad |1\rangle_-$, yielding a $\sqrt{\text{SWAP}}$.

**FIG. 4.** Entanglement using maximally excited dipoles. a) The Ramsey like pulse sequence to achieve a CPHASE gate. <u>1:</u> A pair of atoms are brought together in their ground states. <u>2:</u> A short -pulse brings the (+)-atom to its excited state *iff* it is in the logical-$|1\rangle$ ground state. <u>3:</u> The atoms freely evolve for a time $=\hbar \ /\langle V_{dd}\rangle$. <u>4:</u> A (- ) pulse is applied to the (+)-species, bringing any excited atom to the electronic ground states but with an accumulated minus sign for $|1\rangle_+ \quad |1\rangle_-$. b) The Bloch-sphere representation for this sequence. c) A suitable logical basis to selectively address the (+)-species.

**FIG. 5.** Dipole-dipole figure of merit for spherically symmetric Gaussian wave packets with width $x_0$, normalized to the Lamb-Dicke parameter $= kx_0$, as a function of the normalized separation $\bar{z} = z/x_0$. Maximum $|\ |\ 0.015/^3$ is achieved at $\bar{z} \quad 2.5$.

**FIG. 6.** Schematic of S+P molecular potentials as a function of the internuclear distance *r*. Excitation by a far-off resonance blue detuned laser would normally be dominated by the repulsive potential at the Condon point $R_C$. For well localized but separated atoms (relative coordinate probability distribution sketched here) off-resonant interaction with the attractive potential dominates. The orientation of the dipoles relative to the internuclear-axis is indicated for each potential.

**FIG. A1.** Internal energy levels for two atoms. (a) The bare energy eigenbasis showing degenerate states with one photon excitation. (b) The two atom picture in the rotating frame at field detuning with dipole-dipole splitting of the symmetric and antisymmetric states. The primed states to the right are the dressed atomic states including the sum of the light shift and dipole-dipole potentials.

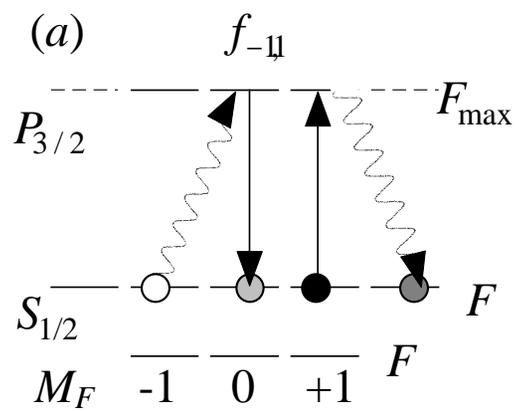

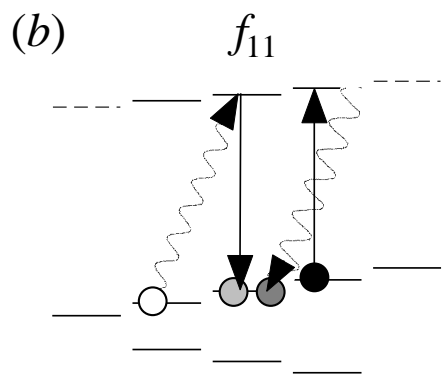

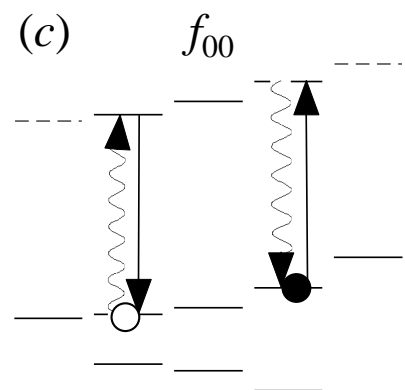

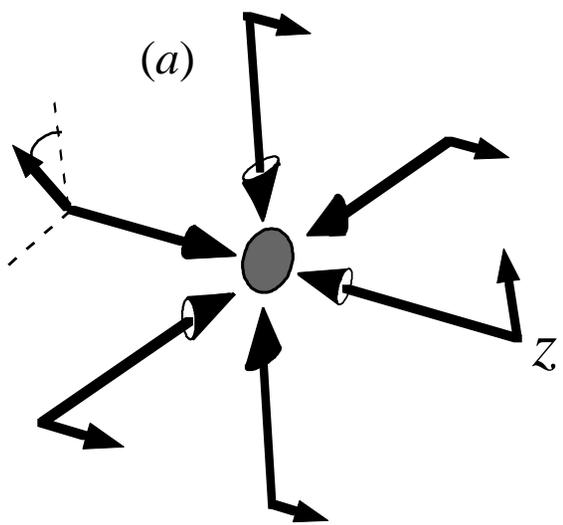 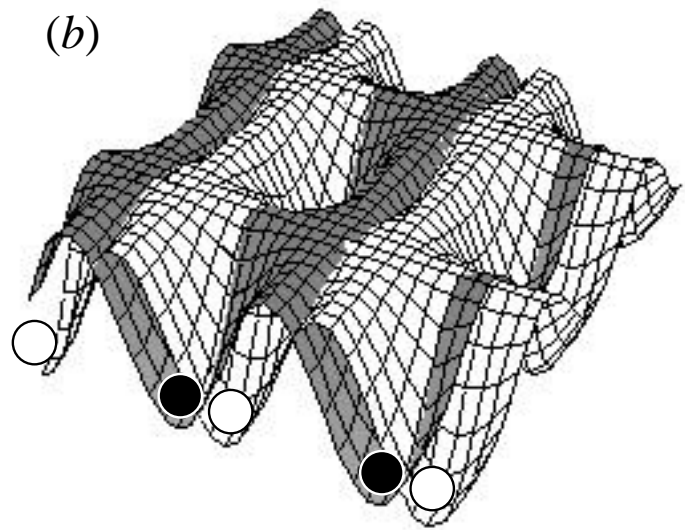

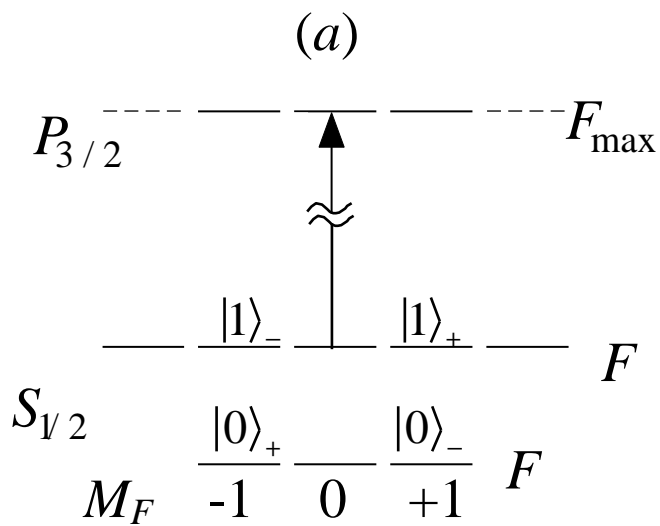

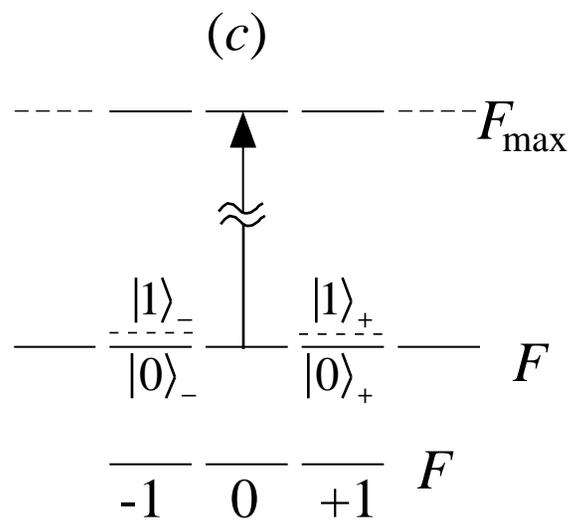

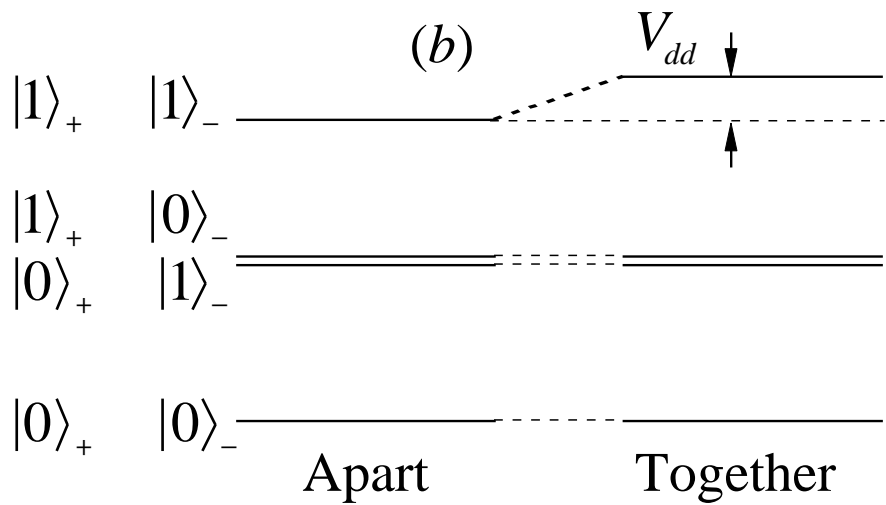

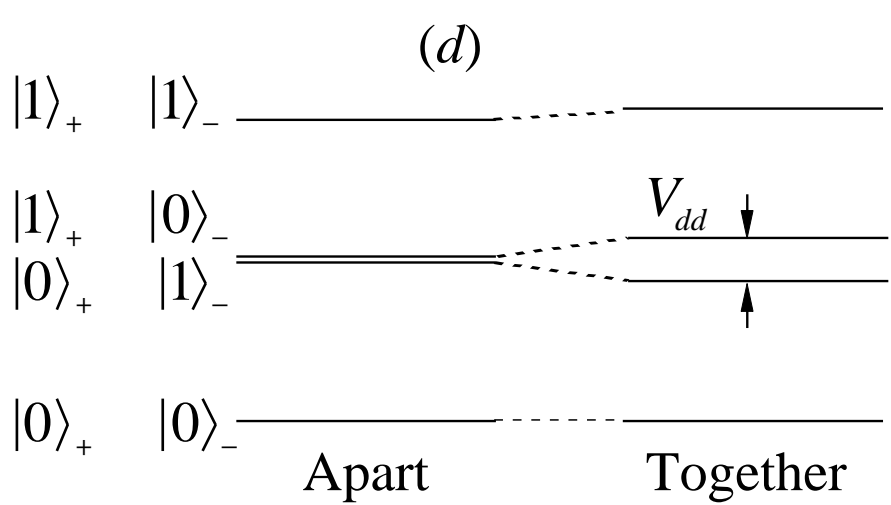

(a)

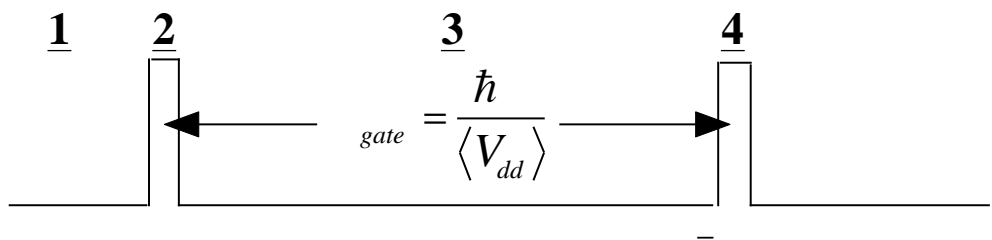

(b)

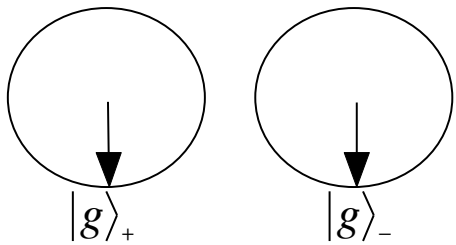
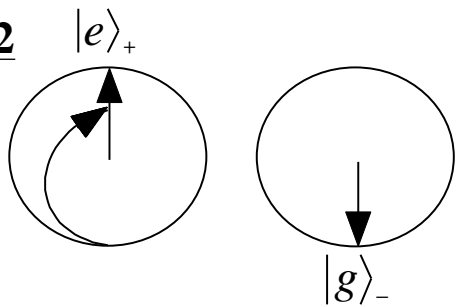

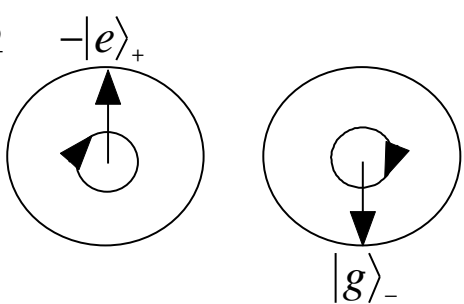
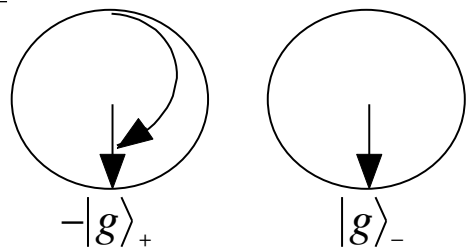

(c)

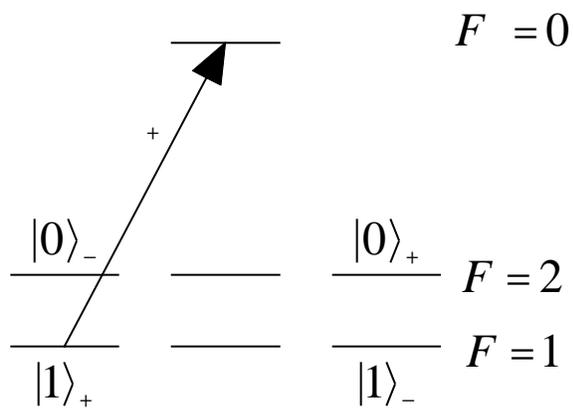

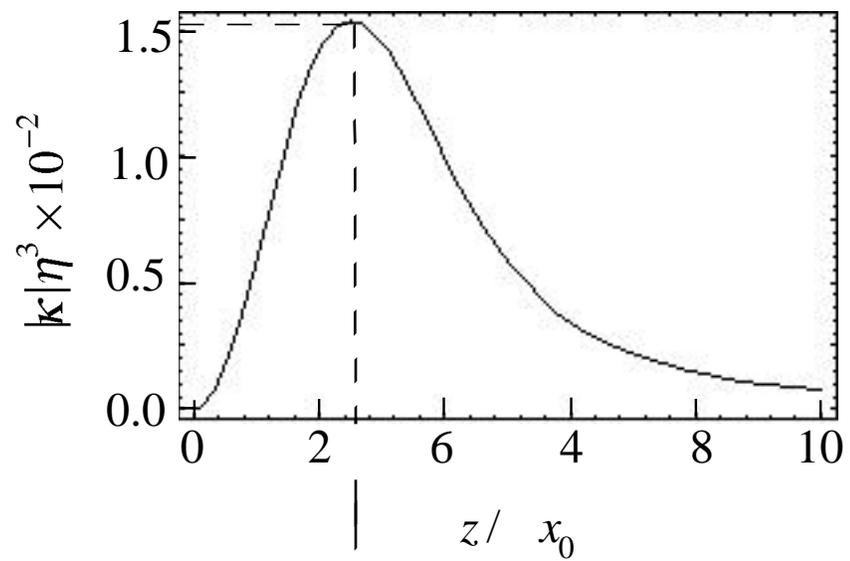

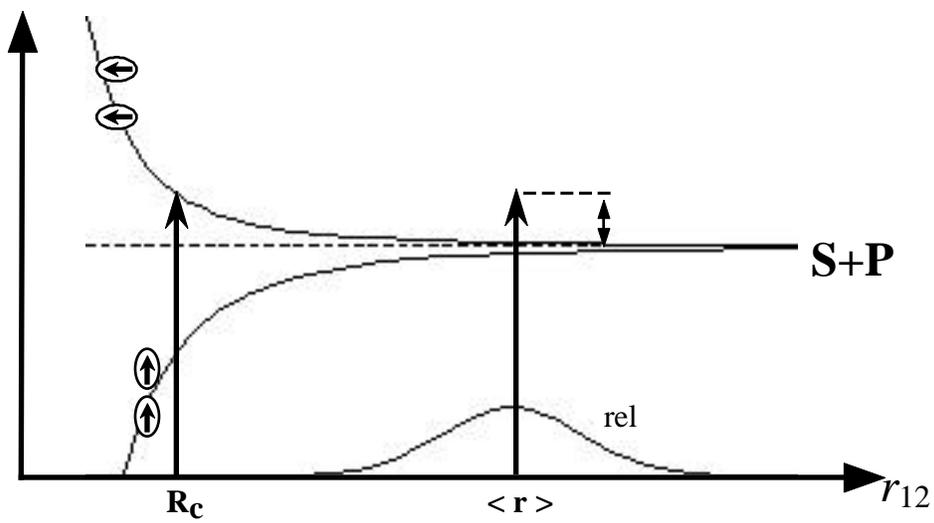

**S+P**

$R_c$        $\langle r \rangle$       $r_{12}$

rel

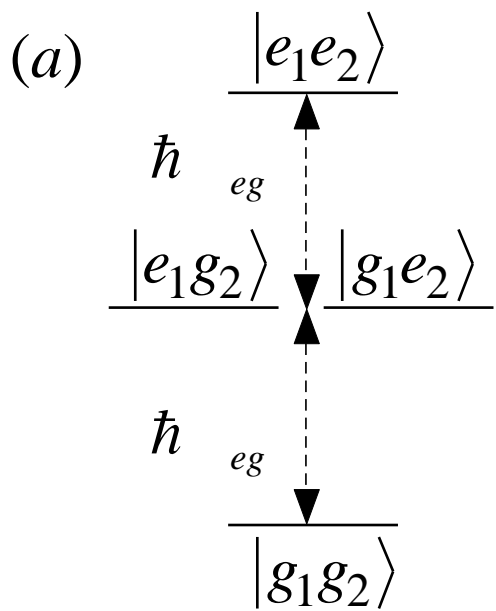

$(a)$

$|e_1 e_2\rangle$

$\hbar\omega_{eg}$

$|e_1 g_2\rangle \quad |g_1 e_2\rangle$

$\hbar\omega_{eg}$

$|g_1 g_2\rangle$

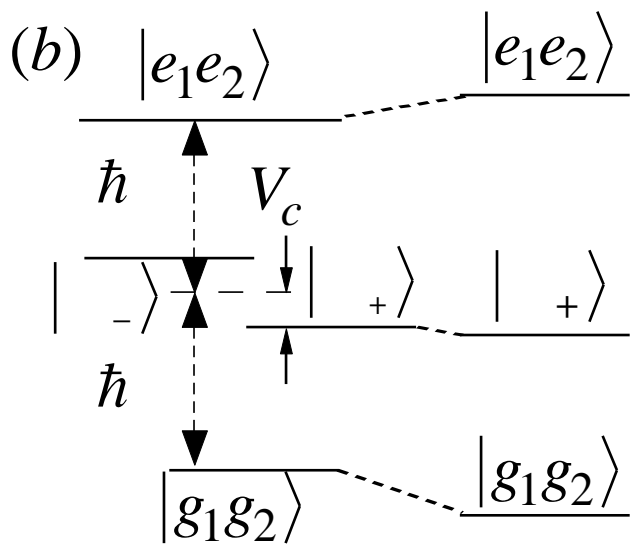

$(b)$

$|e_1 e_2\rangle \qquad |e_1 e_2\rangle$

$\hbar\omega \qquad V_c$

$|-\rangle \quad\quad |+\rangle \qquad |+\rangle$

$\hbar\omega$

$|g_1 g_2\rangle \qquad |g_1 g_2\rangle$